\documentclass[usenatbib,useAMS]{mnras}
\usepackage{natbib}
\usepackage{amsmath}
\usepackage{amssymb}
\usepackage{graphicx}
\usepackage{subfigure}
\usepackage{color}
\usepackage{epsfig}
\usepackage{multirow}
\usepackage{multicol}
\usepackage{dblfloatfix}    
\allowdisplaybreaks[1]

\newcommand{\be}{\begin{equation}}
\newcommand{\ee}{\end{equation}}
\newcommand{\bea}{\begin{eqnarray}}
\newcommand{\eea}{\end{eqnarray}}
\newcommand{\kms}{\,{\rm km \,\, s}^{-1}}

\newcommand{\rsh}{R_{\rm shock}}
\newcommand{\rcool}{R_{\rm cool}}
\newcommand{\msun}{{\rm M}_\odot}

\newcommand{\pc}{\rm pc}
\newcommand{\yr}{\rm yr}
\newcommand{\cc}{{\rm cm}^{-3}}

\newcommand{\sfrarea}{M_\odot~{\rm yr}^{-1} \, {\rm kpc^{-2}}}
\newcommand{\mparea}{\rm M_\odot \, {\rm pc^{-2}}}

\newcommand{\Mcl}{\rm M_{cl}}
\newcommand{\Rcl}{\rm R_{cl}}
\newcommand{\Mgmc}{\rm M_{GMC}}
\newcommand{\eps}{\epsilon_*}
\newcommand{\tsn}{\rm t_{SN}}
\newcommand{\dtsn}{\Delta {\rm t_{SN}}}

\newcommand{\tdnorm}{\rm t_{d,7}}
\newcommand{\Esn}{\rm E_{SN}}
\newcommand{\Psn}{\rm P_{SN}}
\newcommand{\vej}{\rm v_{ej}}
\newcommand{\Mej}{\rm M_{ej}}

\newcommand{\Sigg}{\rm \Sigma_{g}}

\newcommand{\fmix}{\rm f_{mix}}
\newcommand{\fV}{\rm f_{V,0.1}}

\newcommand{\etaE}{\eta_{\rm E}}
\newcommand{\etaM}{\eta_{\rm M}}

\newcommand{\etaP}{\eta_{\rm P}}

\newcommand{\Mhat}{\hat{M}_{\rm hot}}
\newcommand{\Ehat}{\hat{E}_{\rm hot}}
\newcommand{\etacool}{\eta_{\mathrm{cool}}}

\newcommand{\Mdot}{\dot M}
\newcommand{\Edot}{\dot E}
\newcommand{\Edotcool}{\dot{E}_{\rm cool}}

\newcommand{\EdotSN}{\dot{E}_{\rm SN}}
\newcommand{\PdotSN}{\dot{P}_{\rm SN}}
\newcommand{\Edotwind}{\dot{E}_{\rm wind}}
\newcommand{\Mdotwind}{\dot{M}_{\rm wind}}
\newcommand{\Pdotwind}{\dot{P}_{\rm wind}}

\newcommand{\vout}{v_{\rm out}}
\newcommand{\vesc}{v_{\rm esc}}
\newcommand{\vB}{v_\mathcal{B}}

\usepackage[dvipsnames]{xcolor}
\definecolor{halfgray}{gray}{0.55}
\definecolor{webgreen}{rgb}{0,.5,0}
\definecolor{webbrown}{rgb}{.6,0,0}
\definecolor{webpurple}{rgb}{.4,0,.4}
\definecolor{webblue}{rgb}{0,0.6,0.6}

\hypersetup{colorlinks=true,citecolor=webblue}

\title[Clustered Supernovae Drive Powerful Galactic Winds]
{Clustered Supernovae Drive Powerful Galactic Winds after Super-Bubble Breakout}

\author[Fielding, Quataert, \& Martizzi]{Drummond Fielding$^{1}$\thanks{E-mail:drummondfielding@gmail.com}, Eliot Quataert$^{1}$, \& Davide Martizzi$^{1,2,3}$
\\$^{1}$Department of Astronomy and Theoretical Astrophysics Center; University of California, Berkeley, Berkeley, CA 94720, USA
\\$^{2}$Dark Cosmology Centre, Niels Bohr Institute, University of Copenhagen, 2100 Copenhagen, Denmark
\\$^{3}$Department of Astronomy and Astrophysics, University of California, Santa Cruz, CA 95064, USA}

\begin{document}

\pagerange{\pageref{firstpage}--\pageref{lastpage}} \pubyear{2018}
\maketitle
\label{firstpage}

\begin{abstract}

We use three-dimensional hydrodynamic simulations of vertically stratified patches of galactic discs to study how the spatio-temporal clustering of supernovae (SNe) enhances the power of galactic winds.   SNe that are randomly distributed throughout a galactic disc drive inefficient galactic winds because most supernova remnants lose their energy radiatively before breaking out of the disc.   Accounting for the fact that most star formation is clustered alleviates this problem.   Super-bubbles driven by the combined effects of clustered SNe propagate rapidly enough to break out of  galactic discs well before the clusters' SNe stop going off. The radiative losses post-breakout are reduced dramatically and a large fraction ($\gtrsim 0.2$) of the energy released by SNe vents into the halo powering a strong galactic wind.  These energetic winds are capable of providing strong preventative feedback and eject substantial mass from the galaxy with outflow rates on the order of the star formation rate.  The momentum flux in the wind is only of order that injected by the SNe, because the hot gas vents before doing significant work on the surroundings. We show that our conclusions hold for a range of galaxy properties, both in the local Universe (e.g., M82) and at high redshift (e.g., $z \sim 2$ star forming galaxies).   We further show that if the efficiency of forming star clusters increases with increasing gas surface density, as suggested by  theoretical arguments, the condition for star cluster-driven super-bubbles to break out of galactic discs corresponds to a threshold star formation rate surface density for the onset of galactic winds  $\sim 0.03 \, \sfrarea$, of order that observed.

\end{abstract}


\section{Introduction}

Galactic winds are seen to emanate from galaxies across the star forming sequence with outflow rates and velocities correlated with star formation rate \citep{Martin1999,Rubin+14}. These galactic winds are typically invoked to explain the small baryon fraction in low mass galaxies \citep[e.g.,][]{Guo2010}, the observed mass-metallicity relationship of galaxies \citep[e.g.,][]{Tremonti+04}, and the enrichment and heating of the circumgalactic and intergalactic media (CGM and IGM) \citep[e.g.,][]{Oppenheimer2006,Fielding2017a}. The energy injected by supernovae (SNe) -- as well as other forms of stellar feedback such as HII regions, stellar winds, radiation pressure, and/or cosmic rays generated by SNe -- plays a key role in unbinding gas from low-mass galaxies and powering galactic winds \citep[e.g.,][]{Veilleux+05}.

Winds have been observed extensively at many wavelengths probing gas at a broad range of temperatures. The hottest component ($T\gtrsim 10^7$ K) of the winds are measured using their X-ray emission \citep[e.g.,][]{Strickland2009}, intermediate temperature wind material is observed in UV absorption studies \citep[e.g.,][]{Chisholm+16}, and molecular gas is used to map out the coldest phases ($T\lesssim 10^2$ K) \citep[e.g.,][]{Walter+02}.   
However, there is large uncertainty in inferring the amount of mass and energy carried by galactic winds from observations. In general, mass-loading factors (the ratio of the mass outflow rate to the star formation rate) ranging from $\sim 0.1-10$ (see \citealt{Veilleux+05} for review) and order unity energy-loading factors (the ratio of the energy outflow rate to the SN energy injection rate; \citealt{CC85,Strickland2009}) are inferred.  

Cosmological simulations have further demonstrated the critical importance of galactic winds for reproducing the properties of galaxies. Without feedback associated with star formation, or with inefficient feedback, cosmological simulations over-predict the stellar masses of lower mass galaxies. Feedback brings the predicted stellar masses and star formation rates into agreement with observations \citep[e.g.,][]{Hopkins+14}. 
These winds must carry a large fraction of the energy injected by SNe in order to prevent excessive accretion onto galaxies, eject interstellar gas, and thereby affect galaxy evolution.


Numerous numerical studies have sought to understand in detail exactly if/how galactic winds are driven by SNe. In particular, in recent years many groups have adopted a similar approach in which SNe are set off in a stratified medium meant to represent a patch of a galaxy's ISM. These simulations span a wide range in the degree of realism and have been used to address many topics, such as how the galactic wind properties depend on gas surface density \citep[e.g.,][]{Creasey+13} and on the relative scale height of gas and SNe \citep[e.g.,][]{Li+16}, to name just a few. The most realistic simulations of this type include magnetic fields, self-gravity, gravitational collapse induced star-formation, and differential rotation, among other features \citep{Kim+18,Gatto+17}. These simulations have focused on roughly Milky Way like conditions with $\Sigg = 10 \ \mparea$ and have been used to study the equilibrium state of the ISM. More idealized/controlled simulations, such as those presented in this paper, complement the more realistic simulations, by allowing for different aspects of the problem to be isolated and studied in detail. The aggregate results of all of these recent simulations can be summarized as follows: SNe alone (i.e., without also adding cosmic rays and/or radiation pressure) can launch powerful galactic winds (with, say, $\gtrsim$ 10 per cent of the SNe energy in a wind) if the SNe go off in low density regions, which can be achieved by having a highly inhomogeneous ISM due to cooling to $T\lesssim100$ K, having the SNe go off above the gaseous scale height, and/or by having overlapping SN remnants.

It is now well-established that the efficiency of SNe feedback depends sensitively on how they are distributed in time and space, with, for example, SNe distributed randomly in space producing stronger turbulence and winds than SNe correlated with the local density peaks (e.g., \citealt{Gatto2015, Martizzi2016}).  Tightly clustering SNe both spatially and temporally in star clusters further enhances their efficiency in driving turbulence and powering galactic winds (e.g., \citealt{Sharma2014,Fielding2017}).  
Significant clustering is expected since the massive stars that eventually become core-collapse SNe predominantly form in clusters \citep[e.g.,][]{deWit+05} that disperse on $\gtrsim 100$ Myr time scales \citep[e.g.,][]{PortegiesZwart+10}, which is significantly longer than the $\lesssim 30$ Myr lifetime of these stars \citep[e.g.,][]{Leitherer+99}.  
Recent simulations have investigated the evolution of the bubbles blown by clustered SNe (referred to as `super-bubbles') to assess the net momentum injection into the ISM and how much energy remains after radiative losses to power a wind \citep{Gentry2017,Kim2017,Yadav+17}. Although these works disagree on several important fronts, they all indicate that under many conditions the cluster-driven super-bubble will have sufficient time to reach the vertical boundary of the disc in which it is embedded and breakout. These works, however, did not include gravity and a vertically stratified disc and so could not capture the breakout process and post-breakout dynamics. In this paper we extend this line of inquiry by studying both the pre- and post-breakout evolution, and find crucial differences in the energetics in these two different phases.



To start we provide analytic arguments suggesting that, for a wide range of galaxy properties, realistic clustering of SNe in star clusters can lead to a large fraction of the energy produced by SNe venting out into the halo in galactic winds.  We first explain why randomly distributed SNe do not drive strong galactic winds (\S \ref{sec:homog-SN}) and then study the critical role of SNe clustering for producing powerful galactic winds (\S \ref{sec:clustered}).   

After setting the analytic framework we introduce a series of numerical experiments that probe the conditions under which a sizable fraction of the energy liberated by SNe can escape the disc to power galactic winds (\S \ref{sec:sim-results}). We focus our attention on the relatively high gas surface density regime, $\Sigg = 30-300\,\mparea$, appropriate for vigorously star forming galaxies that are seen to launch powerful winds, but also the regime where cooling losses have the potential to dramatically sap the wind potency. To start, we study how the super-bubble driven by the collective effect of numerous SNe propagates through the ISM while confined within the disc (\S \ref{sec:results-prebreakout}). In the cases where the super-bubble reaches the scale height of the disc and can breakout we show that there is a dramatic decrease in radiative losses and an increase in the amount of mass and energy that are carried by the resulting wind (\S \ref{sec:results-postbreakout}). In these experiments we detonate spatio-temporally clustered SNe in discs of varying surface densities, which are either stratified or unstratified and have either no cooling below $10^4$ K making the ISM homogeneous, or inhomogeneous and multiphase with cooling down to $10^2$ K plus turbulent motions driven externally with $\delta v \approx 10$ km/s. The homogeneous and unstratified simulations are less realistic but allow for clearer analysis and provide a useful benchmark in comparison to the more realistic turbulent and stratified simulations in terms of the dynamics and numerical convergence. 
In \S \ref{sec:discussion}, we summarize our findings, and discuss their observation implications, how they compare to existing works, and how they might be affected by missing physics. Finally, in a series of Appendices, we investigate the dependence of our results on spatial resolution (App. \ref{app:res}) and changes to the turbulent realization (App. \ref{app:turb}).

\section{analytic expectations}\label{sec:analytic}
\subsection{Uniformly Distributed SNe Do Not Drive Strong Galactic Winds}\label{sec:homog-SN}

We begin by explaining analytically why SNe that are relatively uniformly distributed throughout a galaxy do not drive efficient winds.   To do so, we 
consider a gas disc with surface density $\Sigg$ and scale-height $h$.   The Kenicutt-Schmidt relation implies a star formation rate surface density of $\dot \Sigma_* \simeq 0.07 \, \left(\Sigg/100 \, \mparea \right)^2 \, \sfrarea$ \citep{Thompson+05,OstrikerShetty2011}.   We assume that for each 100 $M_\odot$ stars formed ($\equiv m_*$) there is a core-collapse SNe.   The resulting SNe rate per unit volume is thus
\be
\dot n_{\rm SNe} \simeq 3 \times 10^{-3} \left(\frac{\Sigma_g}{100 \, \mparea}\right)^2 \, \left(\frac{100 \, {\rm pc}}{h}\right) \,  \frac{\rm SNe}{\rm yr \, kpc^3}.
\label{eq:snrate-homog}
\ee
The ability of the SN remnants to overlap before cooling saps their energy is determined by the porosity $Q_{\rm cool}~=~4/3 \pi \rcool^3 t_{\rm cool} \dot n_{\rm SNe}$ where $\rcool \sim 21 n^{-0.42} \, \pc$ and $t_{\rm cool} \sim 3 \times 10^4 n^{-0.54} \, \yr$ are the cooling radius and cooling time of a SN remnant (e.g., \citealt{Martizzi2015}), respectively, and $n$ is the ambient gas density in cm$^{-3}$ that a typical SNe goes off in.    This will be less than the mean density of the ISM $\langle n \rangle$ because the ISM is inhomogeneous.  For example, in a medium with a log-normal density distribution, as is typical of super-sonic turbulence in the ISM, half the volume is occupied by gas below a density $\sim 0.06 \langle n \rangle (\mathcal{M}/30)^{-1.2}$ where $\mathcal{M} \gg 1$ is the assumed Mach number of the turbulence \citep{FG+13}.  We will thus take $n \equiv f_{\rm V} \, \langle n \rangle \equiv 0.1 \, \fV \, \langle n \rangle$ as a typical value.   Writing $\langle n \rangle = \Sigma_g/2 h m_p$ we can combine the above results to estimate that
\be
Q_{\rm cool} \sim 10^{-3} \left(\frac{\Sigma_g}{100 \, \mparea}\right)^{0.2} \left(\frac{h}{100 \, \pc}\right)^{0.8} \fV^{-1.8}.
\label{eq:por-homog}
\ee
Note that our equation \ref{eq:por-homog} predicts a porosity that is significantly smaller than equation 2 of \citet{MO1977}.   This is because we evaluate the porosity at the cooling time while they evaluated it at the time SN remnants (SNRs) reach pressure equilibrium with the ambient ISM.   We believe that our criterion is appropriate when assessing the ability of SNRs to overlap prior to cooling and drive an energetically efficient wind of the kind envisioned in canonical SNe-driven wind models (e.g., \citealt{CC85}).  

Equation \ref{eq:por-homog} shows that for conditions typical of galactic discs, SNe that are relatively uniformly distributed fail to overlap prior to the onset of radiative cooling.   Most of the SNe energy is thus lost radiatively and SNe cannot drive strong galactic winds. \citet{Fielding2017} showed this explicitly using global simulations of galactic discs that resolve the majority of the SN remnants in the disc.  They found that the fraction of the SNe energy powering a wind could be explained by an analytic model that considers only SNe going off sufficiently far above the disc midplane that the density has dropped to the point where $\rcool \gtrsim h$; these are the supernova remnants that break out of the disc prior to the onset of strong radiative cooling.  The resulting wind energy flux relative to the SNe energy injection rate, known as the energy loading $\etaE$ (see Table \ref{table:definitions}), is given by (their eq. 3 \& Fig. 4)
\be
\frac{\dot E_{\rm wind}}{\dot E_{\rm SN}} \sim 2 \times 10^{-4} \left(\frac{h}{100 \, \pc}\right)^{-3/2} \left(\frac{\Sigma_g}{100 \, \mparea}\right)^{-1}.
\label{eq:edot-homog}
\ee
Equation \ref{eq:edot-homog} is inconsistent (by orders of magnitude) with observational estimates of the energy flux in galactic winds (including the hot gas portion of the wind in M82, for which {\em Chandra} observations suggest $\dot E_{\rm wind} \sim 0.3-1 \, \dot E_{\rm SNe}$; \citealt{Strickland2009}) and the wind powers needed to explain the inefficiency of low mass galaxy formation.  In the remainder of this paper, we argue analytically and numerically that these problems can be rectified by accounting for the fact that star formation is highly clustered.

\subsection{Clustered SNe}\label{sec:clustered}

Most massive stars form in massive star clusters that in turn form in massive giant molecular clouds (GMCs) \citep{deWit+05}.   Observations and theory suggest that to first approximation massive GMCs and star clusters dominate the star formation rate in galaxies (e.g., \citealt{Murray2010b}).  This is because the GMC mass function (e.g., \citealt{Solomon1987}) and star cluster mass function (e.g., \citealt{McKee1997,McCrady2007}) are generally somewhat flatter than $\propto M^{-2}$, and so the most massive systems contain most of the mass/stars.   Moreover, more massive GMCs probably turn a larger fraction of their mass into stars (because it is  harder for feedback to disrupt more massive GMCs; e.g., \citealt{Murray2010,Grudic+18}).   This strong clustering of massive stars and hence SNe can greatly enhance the efficacy of SNe feedback, leading to much stronger winds than suggested by eq. \ref{eq:edot-homog}.

\subsubsection{Star Cluster Properties}

A plausible model of star clusters relates their mass to that of large-scale gravitationally unstable perturbations in the galactic disc in which they reside.  In this case, star clusters have a characteristic mass 
\be
\label{eq:mcl}
\begin{split}
& \Mcl  \simeq \eps \Mgmc \simeq \eps \pi h^2 \Sigma_g \\ & \simeq 10^5 \, \msun \, \left(\frac{\eps}{0.01}\right) \left(\frac{h}{100 \, \pc}\right)^2 \left(\frac{\Sigma_g}{300 \, \mparea}\right),
\end{split}
\ee
where $\Mgmc \simeq \pi h^2 \Sigma_g$ is the Toomre mass of self-gravitating clumps in a galactic disc and $\eps$ is the star cluster formation efficiency.   In more detail, GMCs are expected to have a power-law distribution of masses with the Toomre mass representing the characteristic maximum mass of the distribution.  

We assume that star clusters have a typical size of $\Rcl \sim 10$ pc, although our results are not sensitive to this choice.  We further assume that the star cluster can be modeled as a simple stellar population so that the SNe rate is roughly constant for $t \lesssim \tsn \equiv 30$ Myr \citep{Leitherer+99}.  As a result the SNe rate per unit volume associated with (or bound to) a single star cluster is 
\be
\dot n_{\rm SNe} \simeq \frac{ 3 \, \Mcl}{4 \pi m_* \tsn \Rcl^3} \simeq 8 \left(\frac{\Mcl}{10^5 \, \msun}\right) \left(\frac{\Rcl}{10 \, {\rm pc}}\right)^{-3} \frac{\rm SNe}{\rm yr \, kpc^3}
\label{eq:snrate-cl}
\ee
The enhanced efficiency of SNe feedback is evident comparing equations \ref{eq:snrate-homog} and \ref{eq:snrate-cl}:  the SNe rate per unit volume is larger in the location of a star cluster by a factor of $\sim 2 \times 10^3$!    

\subsubsection{Overlap of SNRs}\label{sec:clustered_overlap}

For sufficiently massive clusters the SNe associated with an individual star cluster generically overlap and thus collectively power a coherent `wind bubble' from the cluster.  
To see this, we note that for an individual SNR, the timescale over which the SNR reaches pressure equilibrium and/or mixes with the ambient ISM is $t_{\rm PE} \sim 2 \times 10^6 \, n^{-0.4} \, (\delta v/10\,\kms)^{-1.4} \, $yrs and the radius of the SNR at that time is $R_{\rm PE} \sim 70 \, n^{-0.12} (\delta v/10\,\kms)^{-1.4} \, \pc$  \citep{Cioffi+88}, where the ambient medium pressure is assumed to be $P = \rho \delta v^2$.  
We assume that because the ISM is turbulent, the timescale $\sim t_{\rm PE}$ is of order the timescale on which the ambient ISM conditions revert to what they were prior to the SNe.    For comparison, the time between SNe in a given cluster is $\Delta t_{\rm SN} = \tsn (m_*/\Mcl)$.   The cluster's SNe can only drive a coherent bubble if $\Delta t_{SN} \ll t_{\rm PE}$, which requires
\be
\Mcl \gg 1500  \, n^{0.4} \, \left(\frac{\delta v}{10\,\kms}\right)^{1.4} \ \msun.
\label{eq:Mcl-min}
\ee
When equation \ref{eq:Mcl-min} is not satisifed, each individual SNR is effectively isolated and the results of \S \ref{sec:homog-SN} are likely to be applicable.   By contrast, when equation \ref{eq:Mcl-min} is satisfied each SNe produced by the cluster goes off in a medium whose properties are set by the cluster's previous SNe.  Moreover, so long as $\Rcl \ll R_{\rm PE}$, which is easily satisfied, the exact size of the star cluster is not that important to the subsequent dynamics.   In this regime, the star cluster feedback should be modeled as a coherent `wind bubble' driven by the cluster's SNe (\S \ref{sec:bubble}).   In fact, because $R_{\rm PE}$ is significantly larger than the size of massive star clusters, many of the SNe within the GMC that formed at the same time as the cluster likely contribute to driving the bubble, not just those associated with the most massive cluster.     

\subsection{Cluster-driven Super-Bubbles}\label{sec:bubble}

The preceding section shows that for a wide range of cluster properties, a cluster's SNe drive a coherent bubble into the ISM.  Here we review analytically some of the expected properties of these bubbles (e.g., \citealt{Weaver1977,McCray1987,KooMckee92}), approximating the multiple SNe as a constant source of mass and energy.  Most of the results summarized here are not new, but provide useful analytic framework for our numerical results to follow in \S \ref{sec:sim-results}.   One difference relative to the previous analytic literature is that we argue that mixing between the ambient medium and the SNe bubble enhances radiative cooling, making the bubble evolution closer to momentum conserving than energy conserving.  This conclusion is supported by \citet{Kim2017}'s numerical simulations and our numerical results in~\S \ref{sec:sim-results}.

We assume that $\dot N_{\rm SNe} = \Mcl/(m_* \tsn)$, so that 
\be
\Mdot = \beta \Mej \dot N_{\rm SNe} = \beta \Mej \frac{\Mcl}{m_* \tsn}
\ee
\be\label{eq:edot_SN}
\EdotSN = \Esn \dot N_{\rm SNe} = \frac{\Esn \Mcl}{\tsn m_*} 
\ee
The parameter $\beta \gtrsim 1$ characterizes mixing of ambient ISM gas into the hot, shocked SNe ejecta.

One model for the collective effect of the cluster's SNe is the steady state wind model of \citet{CC85}, in which the SNe thermalize their energy and drive a steady wind into the ISM, which is in turn the source driving the super-bubble considered here.   \citet{Sharma2014} showed that for the steady wind model to be  applicable the free expansion radius of an individual SNR must be smaller than the termination shock of the \citet{CC85} wind model.  This only occurs for massive clusters $\gtrsim 10^6 n^{3/13} \msun$.    Nonetheless, so long as equation \ref{eq:Mcl-min} is satisfied, the properties of the cluster-driven super-bubble are not strongly affected by whether or not a steady wind is established.   The reason is that the sound crossing time inside the super-bubble is much shorter than the expansion time of the bubble as a whole and so the density and temperature approach roughly constant values inside the bubble, set by the mass and energy supplied by the SNe, but relatively independent of exactly where the SNe ejecta thermalize their energy.    

\vspace{-0.2cm}
\subsubsection{Cooling of SNe-Driven Super-Bubbles}
\label{sec:cool}

In the absence of radiative losses in the SNe ejecta, the radius of the forward shock associated with the super-bubble propagating into a medium of density $\rho$ is given by $R_s \propto (\dot E/\rho)^{1/5} t^{3/5}$.     Including the constants,
\begin{align}
\label{eq:Rsbub}
&R_s \simeq  \ 760 \,  A_E  \left(\frac{t}{\tsn}\right)^{3/5} \ \pc   \ \  ({\rm t < \tsn})  \\
&\mathrm{where\,\,}
A_E \equiv  \left(\frac{\Mcl}{10^5 \, \msun}  \right)^{1/5} \, \left(\frac{n}{\cc}\right)^{-1/5} \nonumber.
\end{align}
The super-bubble is driven by the pressure of the hot SNe ejecta.  It is the cooling of this ejecta, not the forward shock driven into the ISM, that determines whether the non-radiative evolution assumed in equation \ref{eq:Rsbub} is applicable.   Estimating the cooling of the hot SNe ejecta we find that 
\be
\frac{t_{\rm cool}}{t_{\rm exp}} \simeq \frac{2 \times 10^3}{\beta^{3/2} \, n^{2/3}} \, \left(\frac{\Mcl}{10^5 \, \msun}\right)^{-1/3}  \left(\frac{R_s}{100 \, \pc}\right)^{-1/3}
\label{eq:cool}
\ee
where we have assumed free-free cooling dominates and $t_{\rm exp} = R/v$.
Equation \ref{eq:cool} shows that the cooling of the super-bubble is negligible even for densities as high as $n \sim 10^{3} \cc$ unless there is efficient  mixing of the hot SNe ejecta with the surrounding ambient ISM (parameterized here by $\beta \gtrsim 1$ which is larger for higher mass loading of the SNe ejecta). 

The above estimate of $t_{\rm cool}$ in the SNe ejecta assumes that the electron and proton temperatures are equal.   Observations of SNe shocks show, however, that this is not the case for high speed shocks \citep{Ghavamian2007}.   The electron-proton Coulomb collision time $t_{\rm ep}$ is actually quite long and may be the rate limiting step in setting the cooling of the shocked SNe ejecta (see \citealt{FG12} for similar physics in the context of bubbles driven by black hole feedback).    
 We find
 \be
 \frac{t_{\rm ep}}{t_{\rm exp}} \simeq \frac{6000 \, f_e^{3/2}}{\beta^{5/2} n^{2/3}} \left(\frac{\Mcl}{10^5 \, \msun}\right)^{-1/3}  \left(\frac{R_s}{100 \, \pc}\right)^{-1/3},
 \label{eq:tep}
 \ee
where we have assumed $T_e = f_e T_p = 1.3 \times 10^9 f_e/\beta$ K.     So long as $t_{\rm ep} \gtrsim t_{\rm exp}$, most of the thermal energy of the bubble will remain locked in the protons which cannot radiate efficiently.   Equation \ref{eq:tep} demonstrates that if SNe-driven super-bubbles undergo a significant energy conserving phase, properly modeling the cooling of that phase requires taking into account $T_e \ne T_p$ at SNe shocks.   However, since $t_{\rm ep}$ (eq. \ref{eq:tep}) $\lesssim t_{\rm cool}$ (eq. \ref{eq:cool}) for all realizable parameters ($f_e \lesssim 1$ and  $\beta \gtrsim 1$), and they scale the same with $n$ and $R_s$, it is unlikely that electron-proton thermalization will change the bubble cooling by more than order unity.  It is thus unlikely to be dynamically important even though the absence of electron-proton equilibration is important for interpreting observations of young SNe remnants.    Moreover, as with  free-free cooling, the Coulomb coupling timescale is very sensitive to the mixing of the SNe ejecta with the ambient ISM, with $t_{\rm ep}/t_{\rm exp} \propto \beta^{-5/2}$.   

There are two potential mechanisms that generate mixing between the SNe ejecta and the ambient medium:   the Rayleigh-Taylor instability and the fact that the ambient medium is inhomogeneous.   The contact discontinuity between the (denser) shocked ambient medium and the (less dense) shocked SNe ejecta is formally Rayleigh-Taylor stable as the bubble shock and contact discontinuity decelerate into the surrounding ISM.  However, each individual SNR goes through a Rayleigh-Taylor unstable phase as it transitions from free-expansion to the Sedov-Taylor phase. To model the mixing induced by the Rayleigh-Taylor instability it is thus critical to separately resolve each individual SNR, rather than treat the SNe as a source of uniform energy and mass injection as is often done.   

Independent of the Rayleigh-Taylor instability, a second source of mixing is determined by the multiphase structure of the ambient ISM, i.e., the extent to which dense clouds from the ISM penetrate into the SNe ejecta (e.g., \citealt{Kim2017}).   To quantify the importance of mixing increasing the density and cooling rate of the SNe ejecta, we note that the rate at which the ambient medium is swept-up by the forward shock is $\dot M_s = 4 \pi R_s^2 \rho v_s$.  If we assume that a fraction $\fmix \leq 1$ of this  becomes mixed into the SNe ejecta via the combined action of the Rayleigh-Taylor instability and the inhomogeneous ambient medium, we find that the effective $\beta$ of the ejecta is
\be
\beta_{\rm mix} \simeq 1.7 \times 10^3 \, \fmix \, n^{2/3} \,  \left(\frac{\Mcl}{10^5 \, \msun}\right)^{-2/3}  \left(\frac{R_s}{100 \, \pc}\right)^{4/3}.
\label{eq:beta-mix}
\ee
For $\fmix \sim 1$, this is enormous and is sufficient to increase the density and decrease the temperature of the SNe ejecta to $\sim 10^6$ K at which point Coulomb coupling and radiative cooling are far more efficient.     Re-evaluating the cooling of the ejecta given this new density and temperature we find
\be
\frac{t_{\rm cool}}{t_{\rm exp}} \simeq 10^{-2} \, \frac{\fmix^{-2} n^{-2}}{\Lambda_{-22}} \, \left(\frac{\Mcl}{10^5 \, \msun}\right) \left(\frac{R_s}{100 \, \pc}\right)^{-3}
\label{eq:cool2}
\ee
where the cooling function is given by $\Lambda = 10^{-22} \Lambda_{-22}$ ergs cm$^3$ s$^{-1}$.   \citet{Kim2017}'s simulations of super-bubble evolution in a multiphase ISM with $\langle n \rangle \sim 0.1-10$ cm$^{-3}$ suggest that mixing of the SNe ejecta with ambient ISM gas is relatively efficient, so that the rapid cooling implied by equation~\ref{eq:cool2} is probably appropriate regime.  We find the same in our simulations in the early phase of bubble evolution, prior to the bubble breaking out of the galactic disc (see \S \ref{sec:results-prebreakout}).   As we shall show, however, radiative losses become much less significant after breakout (see \S \ref{sec:results-postbreakout}).

\subsubsection{Momentum Conserving Bubbles}
\label{sec:mom}
When radiative cooling saps the bubble of much of its energy, we can approximate the bubble evolution as momentum conserving, with $R_s \propto (\dot P/n)^{1/4} t^{1/2}$ where $\dot P$ is the momentum per unit time supplied by the star cluster.    This implies
\begin{align}
\label{eq:Rsbub-mom}
&R_s \simeq  \ 650 \,  A_P \left(\frac{t}{\tsn}\right)^{1/2} \ \pc   \ \  ({\rm t < \tsn}) \\
&\mathrm{where\,\,}
A_P \equiv \left(\frac{\Mcl}{10^5 \, \msun}  
\frac{\Psn}{3 \times 10^5 \, {\rm km \, s^{-1} \, M_\odot}}\right)^{1/4} \left(\frac{n}{\cc}\right)^{-1/4} \nonumber
\end{align}
and $\Psn$ is the momentum of a typical SNe at the end of the Sedov-Taylor phase, which is only a weak function $\propto n^{-1/7}$ of the density of the medium into which the SNe goes off  (e.g., \citealt{Cioffi+88}).   Equation \ref{eq:Rsbub-mom} also implies that the speed of the forward shock is
\be 
\label{eq:vs-mom}
\begin{split}
v_s \simeq & \ 70 \, n^{-1/2} \left(\frac{\Mcl}{10^5 \, \msun}  \right)^{1/2} \left(\frac{R_s}{100 \, \pc}\right)^{-1} \\ & \times \left(\frac{\Psn}{3 \times 10^5 \, {\rm km \, s^{-1} \, M_\odot}}\right)^{1/2}  \kms.
\end{split}
\ee

\subsection{Application to Galactic discs}\label{sec:discs}

We now evaluate the previous results in the context of galactic discs to determine when the bubble driven by the star cluster will breakout of the disc, potentially contributing to a galaxy-scale outflow.  We define breakout to be satisfied if the bubble reaches $R_s \sim h$ with $v_s \gtrsim \delta v$ for $t < \tsn$ (e.g., \citealt{McCray1987, Koo1992}).    If this is the case the majority of the cluster's SNe go off {\em after} the bubble has broken out of the disc.   This removes the pressure confining the late-time SNe and they are likely to freely expand out into the halo, contributing a large fraction of their energy to a galactic wind.   In \S \ref{sec:results-postbreakout} we demonstrate explicitly using numerical simulations that $\dot E_{\rm wind} \sim \dot E_{\rm SNe}$ post breakout because most of the cluster's SNe can vent into the halo. Thus determining whether or not star cluster driven bubbles can breakout out of galactic discs is critical for understanding the efficiency of galactic winds driven by SNe.   

We consider a gas disc of surface density $\Sigg$ in a spherical potential with circular velocity $v_c$.   The scale-height of the disc is given by $h/r \sim \delta v/v_c$, i.e.,
\be
h \sim 100 \ \delta v_{10} \, \tdnorm \ \pc
\label{eq:h}
\ee
where we have defined $\delta v_{10} = \delta v/10 \kms$ and $\tdnorm = (r/v_c)/(10^ 7 {\rm yr})$.
We assume that the interstellar medium has density
\be
n = f_V \langle n \rangle \simeq 20 \, \left(\frac{f_V}{\delta v_{10} \, \tdnorm}\right) \, \left(\frac{\Sigg}{100 \, \mparea}\right)  \, \cc.
\label{eq:n}
\ee
In contrast to \S \ref{sec:homog-SN}, we do not necessarily assume $f_V \ll 1$ in what follows, even though this is appropriate for the median conditions in the ISM.   The reason is that the mass mixed into the SNe ejecta (and thus the bulk of the overlying ISM) must itself be removed in order for the hot gas produced by later SNe to vent.

To quantify the likelihood of breakout, we assume momentum conserving evolution prior to breakout, evaluate $\Mcl$ using equation \ref{eq:mcl} (scaling $\eps = 0.1 \epsilon_{*,0.1}$), $h$ using equation \ref{eq:h}, and $n$ using equation \ref{eq:n}, to find
\be
\frac{R_s(t=\tsn)}{h} \simeq  \ 4 \, \epsilon_{*,0.1}^{1/4} \, f_V^{-1/4} \, \delta v_{10}^{-1/4} \, \tdnorm^{-1/4}
\label{eq:Resc}
\ee
\be
v_s(R_s = h) \simeq  \ 30 \   \epsilon_{*,0.1}^{1/2} \, f_V^{-1/2} \, \delta v_{10}^{1/2} \, \tdnorm^{1/2} \ \kms.
\label{eq:vesc1}
\ee
Equations \ref{eq:Resc} and \ref{eq:vesc1} show that the most stringent constraint is typically whether the bubble reaches pressure equilibrium with $v_s \sim \delta v$ prior to breakout.      The condition $R_s(t = \tsn) \gtrsim h$ is comparatively easy to satisfy.    Put another way, if star-cluster driven bubbles breakout of galactic discs, they do so quickly, on a timescale $\ll \tsn$, so that most of the SNe associated with the cluster vent at late times contributing their energy to a galactic wind.

The above results can be derived even more simply by asking at what time ${\rm t_{breakout}}$ and speed can a thrust $\dot P$ move the ambient ISM mass of $\pi \Sigma_g h^2$, neglecting gravity or pressure confinement.   The results can be expressed in terms of two basic velocities in the problem.   The first is $h/\tsn \sim 3 \, (h/100 \pc) \, \kms$, the speed to move a distance of order the scale-height before the cessation of the cluster's SNe. The second is $\eps \Psn/m_* \sim 300 \, \epsilon_{*,0.1} \, \kms$, the speed set by the terminal momentum of SNe, scaled by the cluster formation efficiency.    Expressed in these terms, we find
\be
\frac{\rm t_{breakout}}{\tsn} \simeq \left(\frac{h}{\tsn} \, \frac{m_*}{\eps \Psn}\right)^{1/2} \ \ \ {\rm and}
\label{eq:tesc}
\ee
\be\label{eq:vesc}
v_s(R_s = h) \simeq \left(\frac{\eps \, \Psn}{m_*} \, \frac{h}{\tsn}\right)^{1/2},
\ee
which is valuable because it also reveals more directly the dependence on the SNe/IMF properties $\Psn$, $\tsn$, and $m_*$.

\section{Numerical Simulations}
The analytic arguments in the preceding section demonstrate that sufficiently clustered SNe can inflate a bubble 100s of pc in size well before $\tsn$, even in the conservative, momentum-driven limit. This means that in many cases the bubble will be able to reach $h$, the scale height of the galactic disc and breakout. Post-breakout the dynamics and energetics can change dramatically, which will have major implications for the properties of the resulting galactic wind.

We performed a set of controlled simulations to test the conditions under which a large fraction of the SNe energy can escape the disc to power galactic winds. We specifically look at the change in energetics prior to and following breakout, which in turn determines the degree of mass and energy loading of the wind. Our simulations are performed with the Eulerian hydrodynamics code \textsc{athena++}\footnote{\url{https://princetonuniversity.github.io/athena/}} (Stone et al. in prep), which is a recent rewrite of \textsc{athena} \citep{Stone+08}. We adopt a $\gamma = 5/3$ equation of state and evolve the standard hydrodynamics equations with source terms to include optically thin cooling and photoelectric heating, energy and momentum injection from SNe and from externally driven turbulence.  We do not add any explicit thermal or viscous diffusion. 

We study the bubble evolution and breakout process with four types of simulations that are either homogeneous or turbulent, and vertically stratified or unstratified (including an external gravitational potential or not). The turbulent stratified simulations are the most realistic and are our main focus. The homogeneous simulations have no radiative cooling below $10^4$ K and no photoelectric heating, so, with the exception of the bubble material, the ISM is single phase. On the other hand, in the turbulent simulations we allow the gas to cool down to $10^2$ K and include photoelectric heating. The turbulent energy injected prevents the ambient ISM from forming a razor thin disc when gravity is included.

Table \ref{table:sims} summarizes the conditions considered in our simulations.

\subsection{Numerical Method}
\subsubsection{Cooling and Heating}

The energy source term -- the net change in energy per unit time per unit volume -- is given by 
\be\label{eq:cool_def}
\dot{E}_{\rm cool - heat} = - n_H^2 \Lambda(T) + n_H \Gamma. 
\ee
The cooling curve $\Lambda(T)$ we use was made by combining the $T>10^4$ K collisional ionization equilibrium cooling curve provided by \cite{Oppenheimer+13} with the $T<10^4$ K cooling curve developed by \cite{Koyama+2002}. We adopt a photoelectric heating rate $\Gamma =10^{-26} \mathrm{erg/s} \, (\langle n_H\rangle / \mathrm{cm}^{-3})$, which scales with the average density as a means to crudely approximate the increase in photoelectric heating in higher density regions where the star formation rates are higher. We intentionally modeled our cooling and heating implementation on what was used by \cite{Kim2017} to facilitate comparisons between our results. All of the gas in our simulations has fixed solar metallicity. In keeping with the idealized nature of our simulations we also keep the mean molecular mass $\mu$ fixed at the value appropriate for a fully ionized plasma at solar metallicity, which means that the temperature of neutral and partially ionized gas ($T\lesssim10^4$ K) is $\lesssim2$ factor of two lower than it would've been with a variable $\mu$ -- this has a negligible effect on the dynamics.

We impose a cooling time constraint on the hydrodynamics time step so that $dt_{\rm hydro}$ is less than or equal to one quarter of the shortest cooling time $t_{\rm cool}$ in the entire domain. This rather stringent requirement ensures that the operator split implementation of cooling and heating properly captures the dynamics.  Comparison simulations run only using the standard CFL constraint on $dt_{\rm hydro}$ yielded qualitatively different results.

\subsubsection{Supernovae Injection}

We inject SNe using the method developed by \cite{Martizzi2015} that determines the amount of thermal and kinetic energy to inject given the spatial resolution and ambient gas properties. This implementation accounts for the cooling and expansion of the SN remnant below the grid scale and is derived from high resolution simulations of individual SN remnants. In practice this sub-grid model works by first calculating the average properties within a small sphere of radius $r_{\rm inj} = 3 \Delta x$ centered on the location of the upcoming SN, then setting all of the state variables within that sphere to their average value plus the additional mass, energy, and momentum from the SNe.  We have used this SNe injection method in studies of the effect of SNe on an unstratified ISM patch \citep{Martizzi2015}, a stratified ISM patch \citep{Martizzi2016}, and on launching winds from global galactic discs \citep{Fielding2017}. In our current simulations the SNe are seeded at random locations within the cluster radius $R_{\rm cl} = 10$ pc. Their temporal spacing is set by the cluster mass $\dtsn = \tsn/(\Mcl/m_\star)$. Because the SNe are tightly clustered in space and time all but the first few SNe go off within the hot, dilute remnant of a previous SN (so long as $\Delta t_{\rm SN} < t_{\rm PE}$; see \S \ref{sec:clustered_overlap}), so the cooling radii are at least an order of magnitude larger than the injection radius $r_{\rm inj}$. The very large cooling radii relative to the spatial resolution ensures that essentially all of $E_{\rm SN} \equiv 10^{51}$ erg is deposited in the surrounding gas. 

\subsubsection{Turbulence and ISM inhomogeneities}

In the turbulent simulations velocities are driven on large scales such that the mass-weighted velocity dispersion $\delta v=~\langle v^2 \rangle_{M}^{1/2} = 10$ km/s  --  consistent with observed ISM velocity dispersions and roughly equal to the sound speed of $10^4$ K gas. The turbulent kinetic energy injection rate is given by $\Edot_{\rm turb} \approx \rho L_{\rm box}^2 \delta v^3$,
where $L_{\rm box}$ is the horizontal box width.  
The turbulence is driven on large scales with power equally distributed between wave numbers of 1 to 4 in units of $2 \pi/L_{\rm box}$. The velocity forcing is restricted spatially to focus the driving to be within $h$. The relative energy input follows $1+\tanh((h-|z|)/0.5\,h)$. A new realization of the driving pattern is generated every $5\times10^{-3}$ crossing-times ($\equiv 2h/\delta v$) and they are time correlated by an Ornstein-Uhlenbeck process with a correlation time of $1/2$ crossing-time \citep{Lynn+12}. The smoothly changing driving pattern ensures that the turbulence does not develop any unphysical standing patterns or outbursts from impulsive changes.\footnote{Interestingly the driving alone is capable of launching a weak wind from the ISM, as was studied by \cite{Sur+16}.}  The turbulence, heating, and cooling in the turbulent simulations lead to a multiphase medium that is closer to what is expected in reality although additional processes such as self-gravity, magnetic fields, viscosity and conduction would likely change the details of the phase structure (\S \ref{sec:missing_phys}). 

The initial conditions for the turbulent simulations are generated by allowing the turbulence and cooling to proceed for 60 Myr -- many turbulent crossing times and cooling times -- prior to the onset of SN explosions. 

\begin{table}
\centering
\caption{Key simulation properties}
\label{table:sims}
\begin{tabular}{l l} 
\hline\hline
\multirow{2}{*}{ISM structure} & turbulent or  homogeneous, and \\
 & stratified (gravity) or unstratified \\ \hline
mean gas densities & \multirow{2}{*}{$\langle n \rangle = 13.9,\ 139\ \mathrm{cm}^{-3}$} \\
\quad (midplane)& \\ \hline
median gas densities & \multirow{2}{*}{$n_{\rm median} =3.6, \ 32\ \mathrm{cm}^{-3}$} \\
\quad (turbulent sims.) & \\ \hline
gas surface density & \multirow{2}{*}{$\Sigg = 30, \ 300 \ \mparea$} \\ 
\quad (stratified sims.) & \\ \hline 
escape speed to top of box & \multirow{2}{*}{$94 \kms$} \\
\quad (stratified sims.) & \\ \hline 
star cluster masses & $\Mcl = 10^4 - 10^6 \ \msun$ \\ \hline 
$\epsilon_\star \equiv \Mcl / \pi h^2 \Sigg$ & $\sim 0.003 - 0.1 $ \\ \hline
\end{tabular}
\end{table}

\begin{figure*}
\includegraphics[width=7.5in]{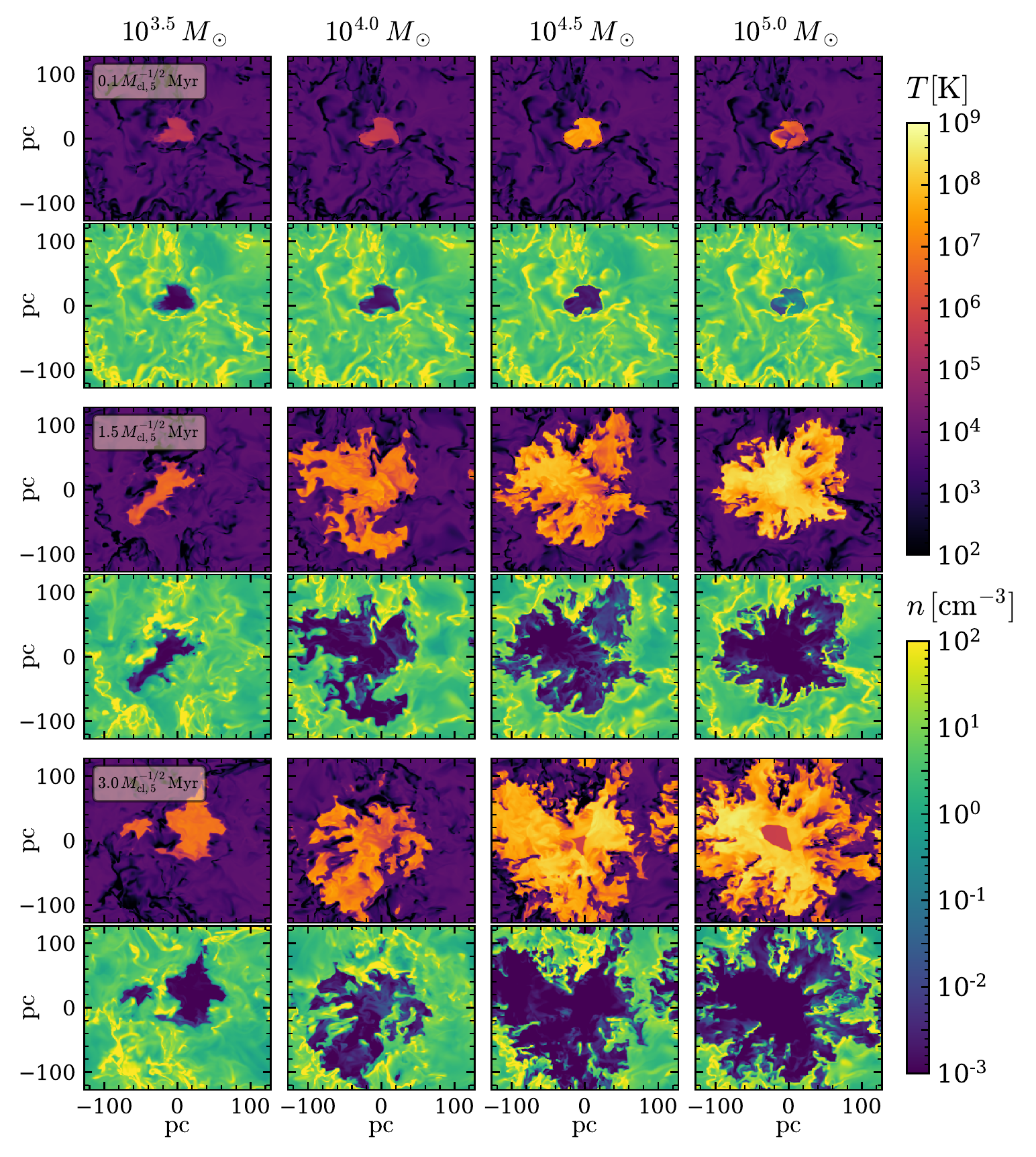}
\caption{Temperature and number density slices through the turbulent unstratified simulations that have a mean density $\langle n \rangle = 13.87\,\cc$. Each column shows the bubble evolution for cluster masses increasing from left to right. Each row shows a snapshot at $t=0.33,1.5$ and $3$ ${\rm M_{cl,5}}^{-1/2}$ Myr, respectively. The times are scaled with $\Mcl^{-1/2}$ to account for the fact that $\rsh\propto t^{1/2} \Mcl^{1/4}$ (see eq. \ref{eq:Rsbub-mom}). The bubbles expand more rapidly in the low density regions causing them to be highly asymmetric. Over dense regions in the ISM penetrate the expanding bubbles and the strong shear forces lead to significant mixing of the ISM and bubble material. }
\label{fig:unstrat}
\end{figure*}

\subsubsection{External Gravity}

By including an external gravitational potential we can study the evolution of bubbles in a vertically stratified medium and what happens after a super-bubble reaches the scale height and breaks out of the disc. In our stratified simulations we adopt a simple gravitational potential that depends only on the height $z$ and represents the vertical component of a spherical potential with circular velocity $v_c$ at a distance of $r$, so $\Phi = \frac{1}{2} (v_c^2 / r^2) z^2$. We adopt $v_c = 175$ km/s and $r=1$ kpc. We neglect the self gravity of the gas $\sim 2 \pi G \Sigma_{\rm gas} z$, which is sub-dominant at most heights up to a gas surface density of $\sim 1000 \, M_\odot$/pc$^2$. Although less realistic, studying the super-bubble evolution without gravity and stratification has the advantage of allowing us to cleanly isolate the pre-breakout phase, so we also present simulations with no external gravity. This has the added benefit of allowing us to connect to the existing numerical simulations of bubble evolution in an unstratified ISM, both inhomogeneous \citep{Kim2017}, and homogeneous \citep{Yadav+17, Gentry2017, Gentry+18}. 

\subsection{Simulation Suite}

\begin{table*}
\centering
\begin{tabular}{lll}
\hline
\hline
Mass & Energy & Momentum \\ \hline
$\Mdot_\star \, ^{\rm a}$ = $ m_* \dot{N}_{\rm SNe} = \dfrac{m_*}{\dtsn}=\dfrac{\Mcl}{\tsn} $     \hspace{0.4 in}
& $\EdotSN \, ^{\rm c}$  = $\Esn \dot N_{\rm SNe} = \dfrac{\Esn \Mcl}{\tsn m_*} = \dfrac{\Esn}{\dtsn }$  
& $\PdotSN \, ^{\rm e}$     = $\Mej\,\vej\,\dot{N}_{\rm SNe}$  \vspace{3pt} \\

$\Mdotwind$ = $\displaystyle{\int} \rho \,(\vec{v}\cdot\hat{z}) dA$ 
& $\Edotwind$ = $\displaystyle{\int \rho \,(\vec{v}\cdot\hat{z}) \left(\dfrac{1}{2}v^2 + \dfrac{\gamma}{\gamma -1}\dfrac{P}{\rho} - \dfrac{1}{2}\vesc^2 \right) dA}$ \hspace{0.4 in}
& $\Pdotwind$ = $\displaystyle{\int} \rho \,(\vec{v}\cdot\hat{z})^2 dA$ \vspace{3pt} \\

$\etaM \, ^{\rm b}$       = $\dfrac{\Mdotwind}{\Mdot_\star}$    
& $\etaE \, ^{\rm d}$    = $\dfrac{\Edotwind}{\EdotSN}$
& $\etaP$                   = $\dfrac{\Pdotwind}{\PdotSN}$ \vspace{3pt} \\

$\Mhat \, ^{\rm b}$       = $\dfrac{\sum\limits_{T>10^{5}\,\mathrm{K}} m_{\rm cell}}{N_{\rm SN} \, m_*}$     
& $\Ehat \, ^{\rm d}$    = $\dfrac{\sum\limits_{T>10^{5}\,\mathrm{K}} E_{\rm cell}}{N_{\rm SN} \, E_{\rm SN}}$ & \vspace{3pt} \\
 & $\etacool \, ^{\rm d}$ = $\dfrac{\EdotSN + \Edot_{\rm turb} - \Edotcool}{\EdotSN}$ & \vspace{3pt} \\
\hline\hline
\end{tabular}
\caption{Definitions of primary quantities used in analysis. 
${}^{\rm a}$ Star formation rate that corresponds to the cluster mass where $m_* = 100 \ \msun$, and $\tsn = 30\,$Myr.
${}^{\rm b}$ $\Mhat$ (eq. \ref{eq:EMhat}) is calculated in the unstratified simulations as a proxy for $\etaM$ in the stratified simulations. 
${}^{\rm c}$ SN energy injection rate where $\Esn = 10^{51}\,$ ergs.
${}^{\rm d}$ $\Ehat$ (eq. \ref{eq:EMhat}) is calculated in the unstratified simulations as a proxy for $\etaE$ in the stratified simulations, and both are compared to $\etacool$, the normalized energy that remains after cooling.   
${}^{\rm e}$ Momentum injection rate by SN where $\Mej = 3\,\msun$, and $\vej = 2.6\times10^3\,\kms$, consistent with $\Esn = 10^{51}\,$ ergs. The momentum per SN is the value injected by our SN model not the asymptotic momentum of an isolated SN in an unstratified medium, which can be $\sim 20-40$ times larger due to work done in the Sedov-Taylor phase. \label{table:definitions}}
\end{table*}

For each of our four types of simulations  --  either turbulent or homogeneous, and stratified or unstratified  --  we adopted two surface densities, $\Sigg$ = 30 and 300 $\mparea$ (really volume densities $\langle n \rangle \approx 13.9$ and $139\,\cc$ for the unstratified simulations). For each surface density we simulated four different cluster masses corresponding to a range in cluster formation efficiencies of $\eps = 0.003-0.1$. The key properties of the simulations are summarized in Table~\ref{table:sims}.

The surface densities we adopt are appropriate for many star forming galaxies and begin to approach the levels seen in the starburst galaxies that launch the most vigorous and readily observable winds. The high surface densities also allow us to study the evolution of the bubbles in the regime where cooling losses have the potential to dramatically reduce their potency. In the unstratified simulations, when we refer to $\Sigg$ we mean that the average density of the simulation is equal to the midplane density of the corresponding stratified simulation, namely of a disc with that $\Sigg$ and a scale height $h=100\,\pc$. The $\Sigg=30$ and $300\,\mparea$ simulations have mean (midplane) densities of $\langle n \rangle=13.9$ and $139\,\cc$, respectively. However, the possibly more relevant density for the turbulent simulations is the median density, or the density of the volume-filling material, since the expanding bubble will follow the path of least resistance. The $\Sigg=30$ and $300\,\mparea$ unstratified turbulent simulations have median densities $n_{\rm median} \approx 3.6$ and $32\,\cc$, respectively, and average  warm gas (above 5000 K) densities of $\langle n_{\rm warm} \rangle \approx 2.6$ and $30.6 \,\cc$.

For the unstratified simulations we adopt a periodic cubical domain that is 256 pc on a side, while the stratified simulations are 512 pc on a side in the horizontal direction and 1080 pc in the vertical direction. The stratified simulations are periodic in the horizontal direction and have modified outflow boundary conditions in the vertical direction that prevent artificial inflows that can arise from a non-zero gravitational acceleration at the boundary. 

Our fiducial spatial resolution is $\Delta x = 2 \,\pc$ for the stratified simulations and 1 pc for the unstratified simulations. In Appendices \ref{app:unstrat_turb_res}, \ref{app:homog}, and \ref{app:strat_turb_res}, we present higher resolution simulations as well and demonstrate that this resolution is sufficient to achieve converged results in most of the quantities of interest.

\section{Simulation Results}\label{sec:sim-results}
We begin by focusing on the evolution of the super-bubble prior to breakout -- paying close attention to the bubble growth and the degree of cooling and mixing in the bubble (\S \ref{sec:results-prebreakout}). This allows us to demonstrate that failure to break out is due to stalling rather than the bubble not having enough time to breakout prior to the cessation of SNe at $t=\tsn$ \citep{Kim2017}. Next, we present the results of the post-breakout evolution, showing, in particular, that once a vent through the ISM is opened the amount of energy lost to cooling drops dramatically and the resulting winds have much higher mass and energy loadings -- on the order of $\sim 0.1-1$ -- than contained in the pre-breakout bubble (\S \ref{sec:results-postbreakout}). For reference, Table \ref{table:definitions} lists and defines the main quantities we focus on in our analysis.

\subsection{Super-Bubble Evolution Within the ISM: Stall or Breakout?} \label{sec:results-prebreakout}

In this section we restrict our attention to the evolution of the super-bubble prior to breakout. To isolate this phase of the evolution we use our unstratified simulations that have no external gravitational potential. The primary quantity we are interested in is the super-bubble radius $r_{\rm bubble}$ to connect to the predictions in \S \ref{sec:analytic} and to assess under what conditions the super-bubble reaches the disc scale height $h$ and can breakout of the disc. We also present the energy and hot gas mass of the bubble as it expands into and mixes with the surrounding ISM. However, as we will show in the next section these quantities are of secondary importance because after breakout the cooling and mixing change dramatically.

Fig. \ref{fig:unstrat} shows temperature and number density slices through the middle of the $\langle n \rangle = 13.9\,\cc$ turbulent simulations with cluster masses ranging from $10^{3.5}$ to $10^5\,\msun$ at three times. The time interval is scaled with $\Mcl^{-1/2}$ to match the expected scaling of a momentum driven bubble (see eq. \ref{eq:Rsbub-mom}).
It only takes the bubble from the most massive cluster a few Myr to reach $\sim100\,\pc$. On the other hand, the least massive cluster's super-bubble is only a few tens of pc in size at this time and has reached pressure equilibrium with the ISM and stalled. 
The super-bubbles expand more rapidly in the low density regions of the ISM and end up enveloping the over dense clumps. These dense clumps experience strong shear forces, which leads to significant mixing. As shown in \S \ref{sec:cool}, the degree of mixing is critical for the bubble evolution since an order unity $\fmix$ can cause the bubble material to cool very rapidly.

\begin{figure}
\includegraphics[width=\columnwidth]{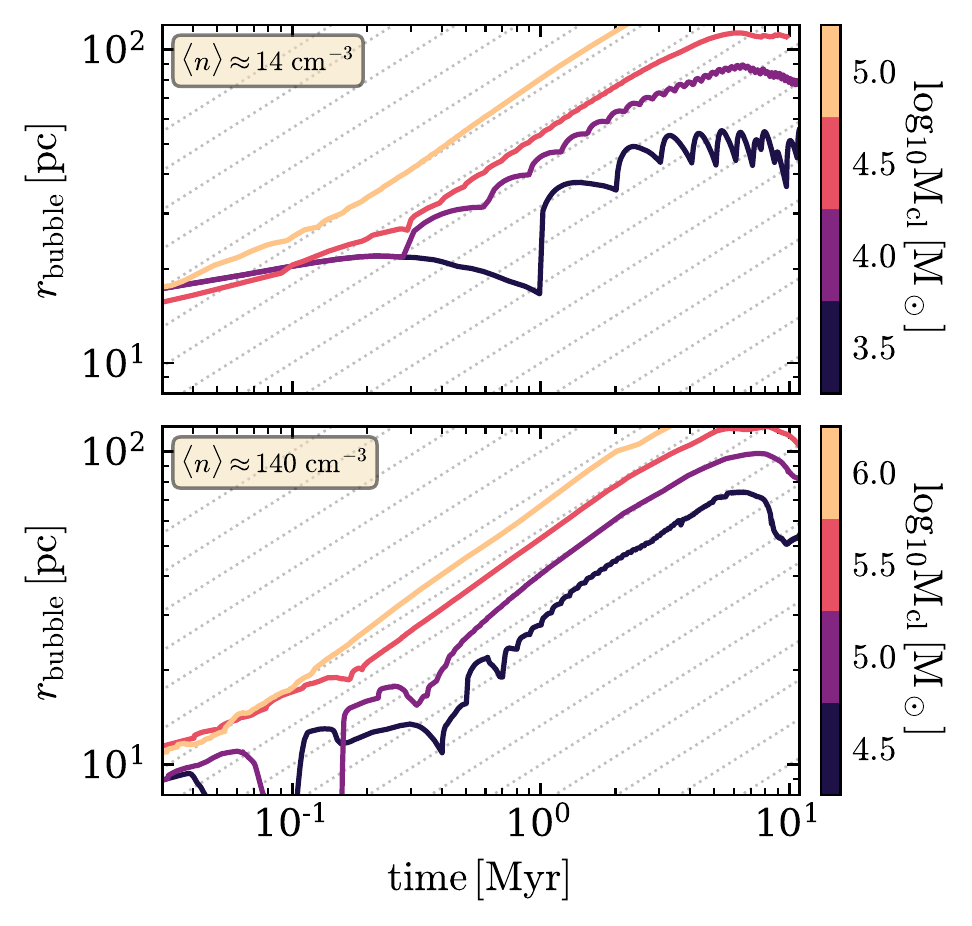}
\caption{The bubble radius evolution in the unstratified turbulent simulations for a range of cluster masses. The top and bottom rows are for the simulations with mean densities $\langle n \rangle=13.9$ and $139\,\cc$, respectively. These correspond to a midplane density appropriate for a galactic disc with $h=100\,\pc$ and $\Sigg=30\,\mparea$ (top), and $\Sigg=300\,\mparea$ (bottom). The dotted grey lines trace $t^{1/2}$. More massive clusters succeed in overcoming the ISM pressure and blow significant bubbles; the radius increases as $t^{1/2}$ indicating that the momentum driven limit applies (see equation \ref{eq:Rsbub-mom}). The bubbles in lower cluster mass simulations stop growing and stall at some $t<\tsn$.}\label{fig:r_sh_evo}
\end{figure}

Before looking at the bubble radius evolution we must first define how we identify it. There are several possible choices for measuring the size of the bubble. In the homogeneous ISM simulations it is straightforward to separate the swept up ISM material from the unperturbed ISM material with a velocity cut since the unperturbed ISM is initially at rest (see \citealt{Kim2017} in which the ISM was inhomogeneous but static, and \citealt{Yadav+17,Gentry+18} whose simulations adopted a purely homogeneous static ISM). However, in the turbulent simulations the ISM is not static so we instead adopt a temperature cut. We classify all gas with $T>10^5$ K as bubble material\footnote{Absent the bubble material no gas is in this temperature range.} and define an effective bubble radius to be 
\be \label{eq:rbubble}
r_{\rm bubble} = \left( \frac{3}{4 \pi} \sum\limits_{T>10^{5}\,\mathrm{K}} \Delta x^3 \right)^{1/3}.
\ee
The analytic predictions for the bubble radius evolution presented in \S \ref{sec:analytic} are technically for the forward shock (shocked interstellar material), so eq. \ref{eq:rbubble} has the potential to miss regions of the ISM that have been swept up and shock heated, but then cooled back down below $10^5$ K. However, the thickness of the cooled swept up shell is very small compared to the radius, so the error this introduces is small.

Fig. \ref{fig:r_sh_evo} shows the growth of the bubble radius in the turbulent simulations for both densities and the full range of cluster masses (star cluster formation efficiencies $\eps$). Since the shock radius scales with $\Mcl/n$ to the 1/5 or 1/4 power in either the energy- or momentum-driven limits (equations \ref{eq:Rsbub} and \ref{eq:Rsbub-mom}), and $\Mcl$ scales linearly with $n$ (equation \ref{eq:mcl}) we do not expect at fixed $\eps$ for there to be a significant dependence on $\langle n \rangle$. This is indeed born out in fig. \ref{fig:r_sh_evo} when comparing different densities. 
The temporal scaling of super-bubble radius shown in Fig. \ref{fig:r_sh_evo} gives us a clue into whether the bubbles are being driven by energy or momentum. Momentum-driven bubbles evolve with $t^{1/2}$ while energy-driven bubbles evolve with $t^{3/5}$. Although the difference in slope is minor, the bubbles predominantly follow the $t^{1/2}$ scaling quite closely (shown with the thin grey lines). This agrees with the findings of \cite{Kim2017} in their similar unstratified inhomogeneous simulations. 

In the context of powering galactic winds the key result in Fig. \ref{fig:r_sh_evo} is that under a broad range of conditions clustered SNe-driven super-bubbles can reach the disc scale height $h$ -- generally on the order of 100 pc -- prior to the cessation of energy injection by SNe at $\tsn$. 
This is true for star cluster formation efficiencies $\eps \gtrsim 0.03 $ (which corresponds to $\Mcl \gtrsim 10^{4.5}$ and $10^{5.5}\,\msun$ for $\Sigg = 30\,\mathrm{and}\,300\,\mparea$, respectively).
For both ISM densities the bubbles driven by the highest $\Mcl$ ($\eps = 0.1$) reach 100 pc in $\sim 2$ Myr and fill the computational volume prior to stalling (we halt the simulations when $r_{\rm bubble} \approx L_{\rm box}/2$ or $t=\tsn$, which ever comes first). The second highest cluster masses ($\eps = 0.03$) reach 100 pc by $\sim3-4$ Myr, and stall soon after. The bubbles powered by the lowest two cluster masses never reach 100 pc, but instead reach pressure equilibrium and stall at $\sim 80$ and 50 pc for $\eps = 0.01$ and 0.003, respectively. In all of the simulations it is stalling rather than running out of time that sets the maximum extent of the hot bubble.
 
The critical $\eps$ that determines whether a bubble will make it to $h$ prior to stalling can be found by equating the shock velocity at the scale height with the turbulent velocity dispersion (see eq. \ref{eq:vesc}). This critical value is given by
\begin{eqnarray} \label{eq:eps_crit}
\epsilon_{\star, {\rm crit}} &=& 0.015 \, \left(\frac{f_V}{0.25}\right) \, \left(\frac{\delta v}{10\kms}\right)^{2} \nonumber \\
&& \quad \left(\frac{h}{100\pc}\right)^{-1} \, \left(\frac{\Psn}{10^5 \msun \, \kms}\right)^{-1},
\end{eqnarray} 
where we have normalized $f_V$ by $n_{\rm median} / \langle n \rangle \approx 0.25$ (see Table \ref{table:sims}). Below this $\epsilon_{\star, {\rm crit}}$ the bubble will never breakout and will instead reach pressure equilibrium with the ISM and stall. 
This analytic prediction is in close agreement with the numerical results. There is, however, a factor of roughly 3 uncertainty in the appropriate value to adopt for the momentum injected per SNe $\Psn$ at the time of breakout \citep{Kim2017}.

The amount of energy and mass contained within the bubble at the time of breakout has been used a proxy for the resulting wind's energy and mass loadings \citep{Kim2017}. These quantities encode how much mixing of the ISM and bubble material has occurred and how much energy has been lost to cooling. Following \cite{Kim2017} we define the following normalized bubble energy and mass:
\be \label{eq:EMhat}
\hat{E}_{\rm hot} = \frac{\sum\limits_{T>10^{5}\,\mathrm{K}} E_{\rm cell}}{N_{\rm SN} \, E_{\rm SN}} \quad \mathrm{and} \quad
\hat{M}_{\rm hot}= \frac{\sum\limits_{T>10^{5}\,\mathrm{K}} m_{\rm cell}}{N_{\rm SN} \, m_{\star}},
\ee
where $N_{\rm SN}$ is the number of SNe that have gone off thus far.
These quantities are analogous to the standard energy and mass loading of the wind -- the outflow rates normalized by the injection rates -- that we define to be
\bea\label{eq:eta_def}
\etaE &=& \frac{\Edotwind}{\EdotSN} = 
\frac{\Mdotwind}{\EdotSN}\left(\frac{1}{2}v^2 + \frac{\gamma}{\gamma -1}\frac{P}{\rho} - \frac{1}{2}\vesc^2 \right) \\
\etaM &=& \frac{\Mdotwind}{\Mdot_\star}\label{eq:etaM_def}
\eea 
with $\Mdot_\star = m_* \dot{N}_{\rm SNe}$. Eq. \ref{eq:EMhat} is appropriate for the unstratified simulations in which there is no wind, while we use equations \ref{eq:eta_def} and \ref{eq:etaM_def} for the stratified simulations.

\begin{figure}
\includegraphics[width=\columnwidth]{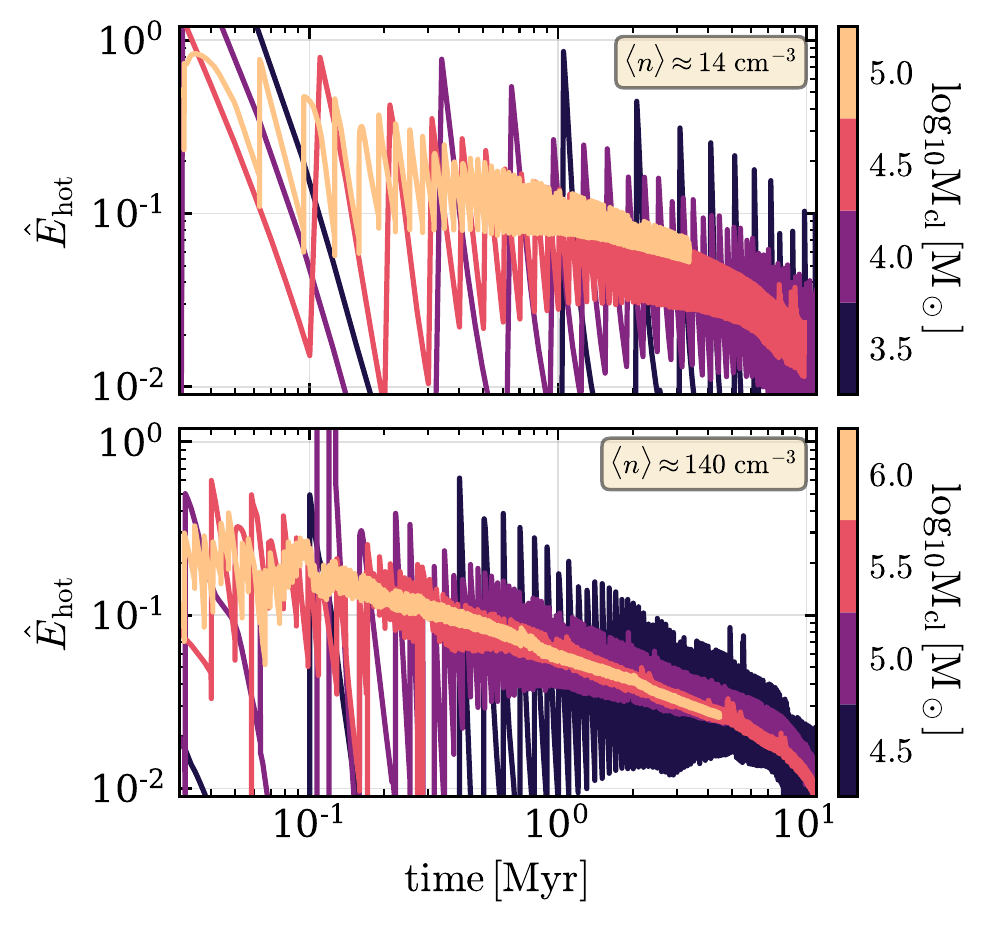}
\caption{$\Ehat$ evolution for the unstratified turbulent simulations for a range of cluster masses. $\Ehat$ quantifies the fraction of the SNe energy retained as thermal energy of the bubble (eq. \ref{eq:EMhat}), and is a proxy for the wind energy loading. The top and bottom rows are for the simulations with mean densities $\langle n \rangle=13.9$ and $139\,\cc$, similar to the $\Sigg=30$ and $\Sigg=300\,\mparea$ stratified simulations, respectively.}\label{fig:Ehat}
\end{figure}

\begin{figure}
\includegraphics[width=\columnwidth]{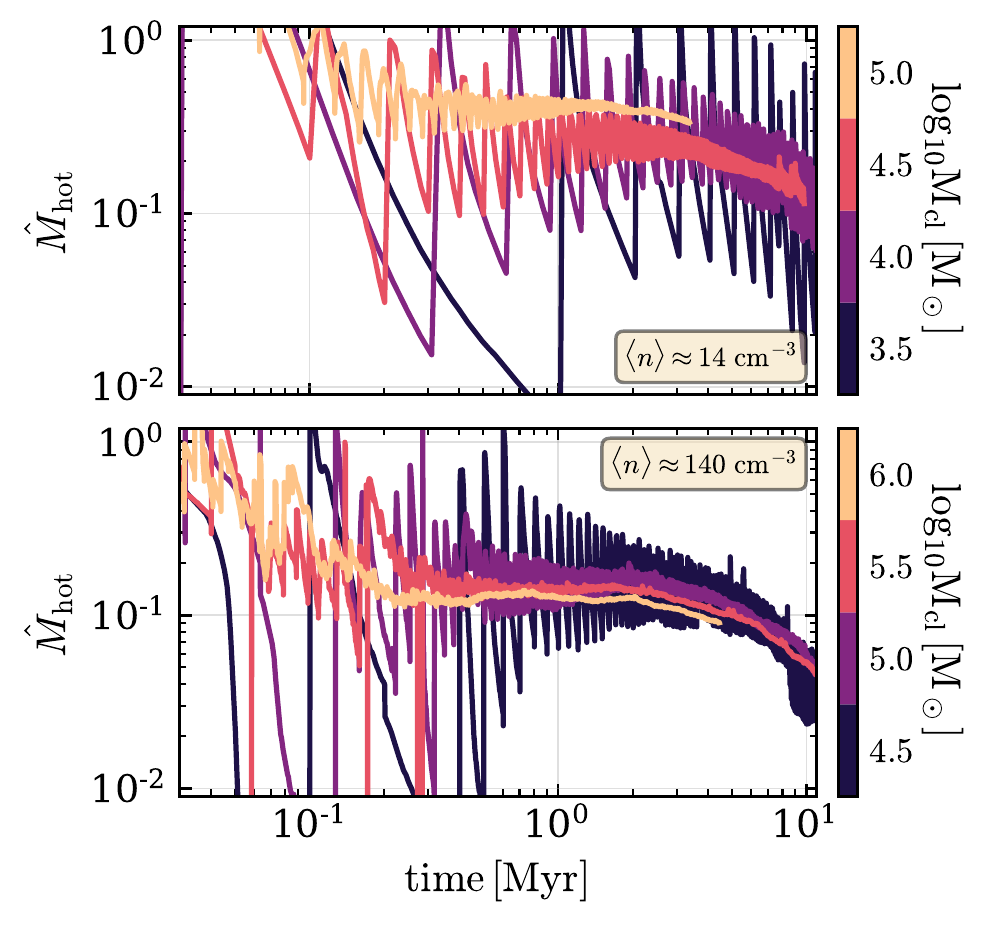}
\caption{$\Mhat$ evolution for the unstratified turbulent simulations for a range of cluster masses. $\Mhat$ is the amount of hot gas relative to the amount of stars formed (eq. \ref{eq:EMhat}), and is a proxy for the wind mass loading. The top and bottom rows are for the simulations with mean densities $\langle n \rangle=13.9$ and $139\,\cc$, similar to the $\Sigg=30$ and $\Sigg=300\,\mparea$ stratified simulations, respectively.}\label{fig:Mhat}
\end{figure}

Figs. \ref{fig:Ehat} and \ref{fig:Mhat} show the evolution of $\Ehat$ and $\Mhat$ for the same simulations in Fig. \ref{fig:r_sh_evo}. Except at very early times well before the bubble can breakout, $\Ehat$ is less than 0.1 and may be as small as 0.01, even for the most massive clusters. Likewise, $\Mhat$ is rarely larger than 0.2. 
As we discuss in Appendix \ref{app:homog} $\Ehat$ and $\Mhat$ are also below 0.1 in the unstratified homogeneous simulations indicating that there is non-negligible mixing even without the large inhomogeneities in the ISM. This degree of mixing is significantly more than found by \cite{Gentry2017} who used a spherically symmetric Lagrangian code capable of resolving the contact discontinuity better than is possible with the Eulerian code and Cartesian grid we used for these simulations. It is, therefore, reasonable to worry that the mixing in our case may be artificial and owing to numerical errors. However, as we show in Appendices \ref{app:unstrat_turb_res} and \ref{app:homog} we find that our both our homogeneous and turbulent simulations are very well converged in $\Ehat$ and $\Mhat$ down to a resolution of $\Delta x = 0.25\,\pc$. We, therefore, believe that to a large extent the ISM-bubble mixing in both the homogeneous and turbulent simulations is real. One source of mixing not captured in 1D codes is that the SNe in our simulations are set off at locations distributed within $10\,\pc$ of each other which leads to complex internal bubble dynamics and asymmetrical acceleration of the shock. Additional mixing can arise because the energy injection within the bubble is not continuous, so the boundary of the bubble experiences impulsive accelerations after each SNe. These accelerations push the less dense bubble material into the more dense shell leading to the development of Rayleigh-Taylor instability, which we discuss in more detail in Appendix \ref{app:RT}. Finally, in the turbulent simulations the cold clumps that penetrate the bubble are ablated due to the strong shear forces they experience. It is important to note that all of these mixing processes are hard to model numerically. Although in Appendix \ref{app:unstrat_turb_res} we show that our results are well converged and do not depend sensitively on our resolution, we caution against over interpreting our findings on these highly mixing dependent quantities to more than a factor of a few level. This is highlighted by the fact that \cite{Kim2017} performed a similar set of numerical experiments using a similar method (albeit with a static inhomogeneous ISM compared to our turbulent ISM, different SNe injection model and equation of state, and different Riemann solvers) and found $\sim3$ times lower values of $\Ehat$ and $\Mhat$. Suffice it to say that at the time of breakout $\Ehat$ and $\Mhat$ are both likely no more than $10^{-1}$ and may be as small as $\lesssim 10^{-2}$.

\cite{Kim2017} argued that $\Ehat$ and $\Mhat$ are representative of the energy and mass loading ($\etaE$ and $\etaM$) of the galactic winds that would result once the bubble breaks out of the disc. 
We now show, however, that the post breakout winds are in fact much more powerful than suggested by Figs. \ref{fig:Ehat} and \ref{fig:Mhat}.

\begin{figure*}
\includegraphics[width=6.9 in]{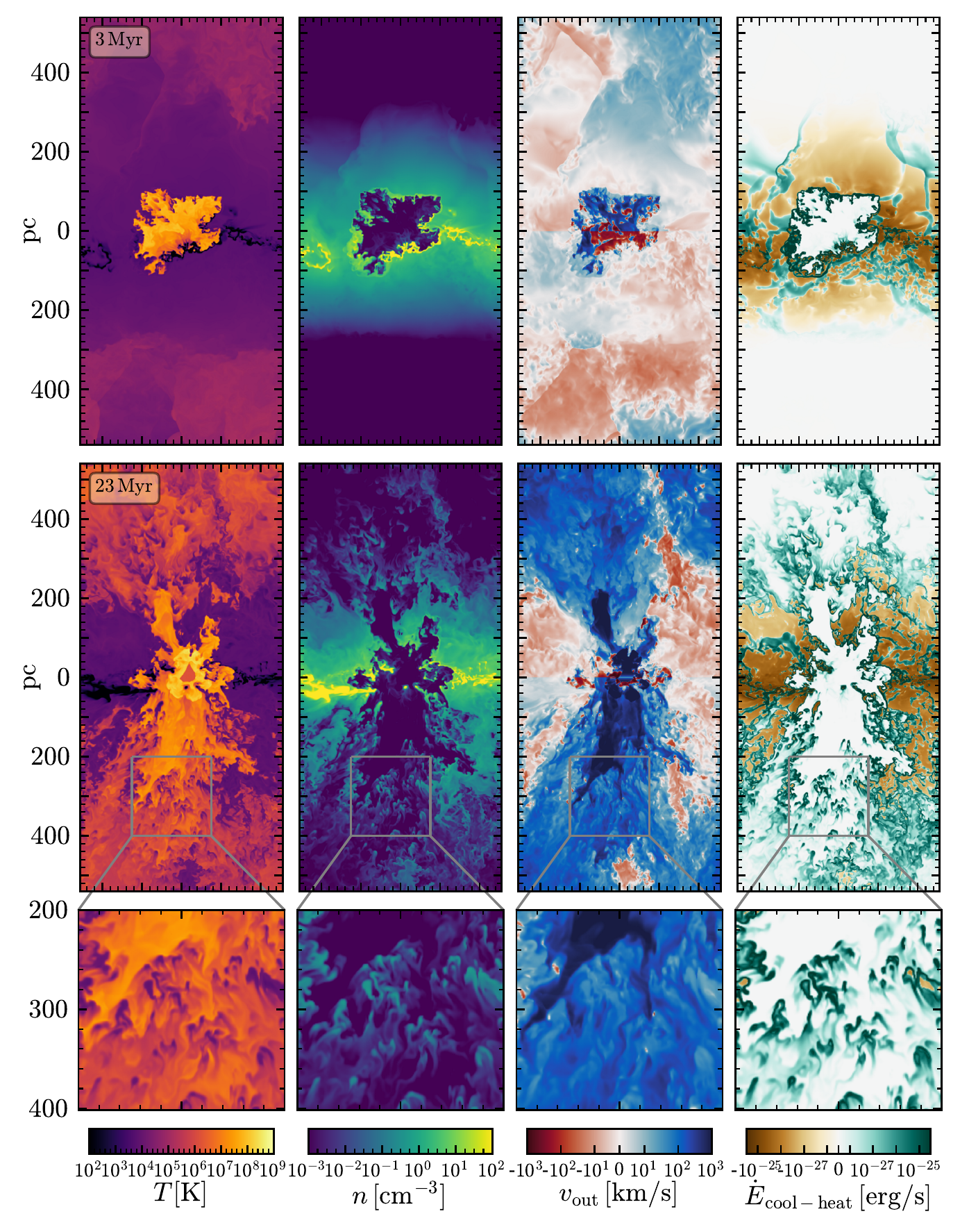}
\caption{Vertical slices through the $\Sigg=30\,\mparea$, $\Mcl=10^{4.5}\,\msun$, $\Delta x = 1$ pc turbulent stratified simulation showing from left to right the temperature, number density, outflowing velocity ($\vec{v}\cdot\hat{z}$, positive means leaving the box), and the cooling/heating rate (positive means losing energy) at $t=3$ Myr, prior to breakout, and near $\tsn$ at $t=23$ Myr, well past the initial breakout, in the top and middle rows, respectively. The bottom rows show zoomed-in patches on a region above the disc that exhibits significant cold cloud entrainment -- these clouds are also growing due to cooling of the enhanced cooling of the hotter medium in their wakes.}
\label{fig:strat}
\end{figure*}

\subsection{Post-breakout Super-Bubble evolution: Wind properties } \label{sec:results-postbreakout}
In the previous section we looked at the properties of super-bubbles while confined within the ISM, prior to reaching the scale height of the disc. We now focus our attention on what happens once the super-bubble pushes its way through the disc and is able to breakout into the surrounding medium. Therefore in this section we primarily focus on the stratified simulations. We begin with a qualitative look at the properties of the super-bubble and post-breakout wind. Then we show that the pre-breakout energetics have little bearing on the post-breakout energetics. Finally, we discuss the wind mass and energy loading for different choices of $\Sigg$ and $\Mcl$ -- highlighting the temperature dependence of the wind loading.

Fig. \ref{fig:strat} shows the state of the $\Sigg = 30 \, \mparea$, $\Mcl = 10^{4.5}\, \msun$ simulation at $t=3$ (top row) and $23$ Myr (middle and bottom rows). From left to right the columns show slices of the temperature $T$, number density $n$, outward velocity $v_{\rm out}\equiv \vec{v}\cdot\hat{z}$, and $\dot{E}_{\rm cool-heat}$ (eq. \ref{eq:cool_def}), respectively. At $t=3$ Myr the bubble is still confined within the disc and there is a large amount of cooling on the boundary of the bubble. At these early times the stratified and unstratified simulations are qualitatively similar. However, post-breakout the properties are entirely different as shown in the $t=23$ Myr panels. Upon reaching the scale height of the disc ($h\approx100\,\pc$) the super-bubble loses the confining pressure in the vertical direction and is able to open a wide `chimney' in the ISM through which it can vent into the surrounding medium without suffering appreciable cooling losses. In the horizontal direction the cavity is continually pressed upon by the thermal and ram pressure of the turbulent ISM. Dense ISM clumps are able to penetrate the cavity wall where they are immediately buffeted, shredded, and/or entrained by the wind. These clumps are responsible for much of the mass loading of the wind and can be seen in different stages of their shredding/entrainment in Fig. \ref{fig:strat} as the cold dense clumps above and below the disc. This shredding and entraining process can be seen in better detail in the zoom-in panels in the bottom row. In some cases, as these clouds are accelerated they grow due to the enhanced cooling of the hot medium in their wakes \citep{Gronke+18}.

Even without the dense ISM clumps the wind would still be appreciably mass loaded, since in the homogeneous ISM stratified simulations (not shown here) the mass loading is greater than pure SNe ejecta loading would imply ($\Mej / m_* = 0.03$ in our case). In the homogeneous simulations the ISM mixing results from the asymmetrical SNe distribution and the strong shear flows on the cavity walls that lead to Kelvin-Helmholtz instabilities. Thus, a combination of effects work in concert to continually mass load the winds. The mass loading can come at a cost to the wind energy. The shredded clouds increase the wind density and decrease the wind temperature, which increases the wind cooling rate -- shown clearly in the middle and lower right panels of fig. \ref{fig:strat}. 

\begin{figure}
\includegraphics[width=\columnwidth]{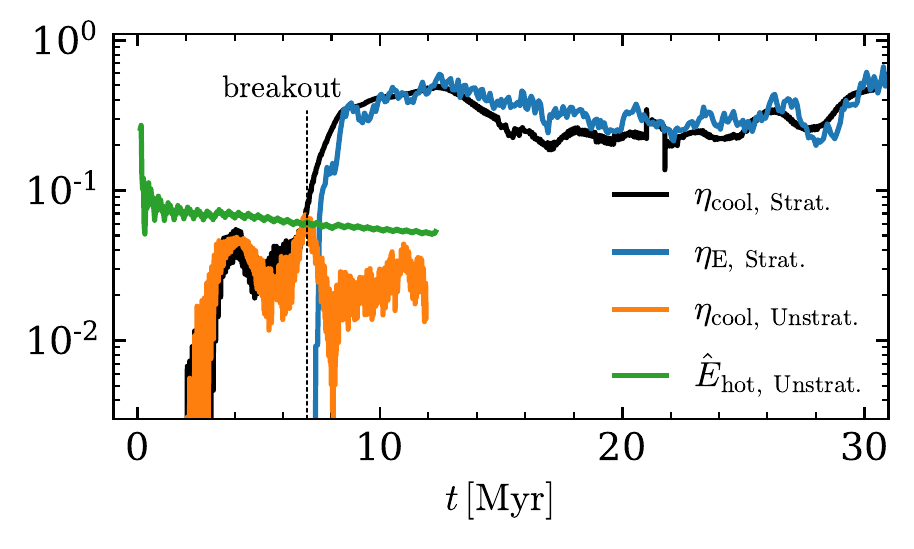}
\caption{Time evolution of $\etacool$ for the stratified (black) and unstratified (orange) $\Sigg = 30 \, \mparea$, $\Mcl = 10^{4.5}\, \msun$ homogeneous simulations, compared with $\etaE$ leaving the top and bottom of the stratified simulations domain (blue line), and $\Ehat$ from the unstratified simulation (green line). The correspondence between the $\etacool$ and $\hat{E}_{\rm hot}$ or $\etaE$ for the unstratified and stratified simulations, respectively, indicates that the injected energy not radiated away goes into the energy of the bubble and the wind. The vertical dashed line at 7 Myr demarcates when the bubble breaks out in the stratified simulations. After this time the $\etacool$ of the stratified and unstratified simulations  differ by an order of magnitude demonstrating that once a channel through the ISM has been opened the energy is able to vent into the lower density surroundings where it experiences significantly less cooling. Most of the SNe go off at $t > 7$ Myr leading to a time average energy loading of $\etaE \gtrsim 0.2$.}
\label{fig:homog_eta_E_leftover}
\end{figure}

We now quantify how much of the energy injected into the ISM is lost to radiative cooling. We demonstrated in the previous section that while the bubble is confined within the disc on the order of 90 to more than 99 per cent of the energy injected by SNe was lost to cooling. We can assess this in the stratified case by measuring the difference between the energy injected and the energy lost to cooling relative to the injected energy SNe energy. We call this quantity $\etacool$ and it represents the energy that is leftover to power the wind:
\be 
\etacool = \frac{\EdotSN + \Edot_{\rm turb} - \Edotcool}{\EdotSN} = 1 + \frac{\Edot_{\rm turb}}{\EdotSN} - \frac{\Edotcool}{\EdotSN}.
\ee
Recall from equation \ref{eq:edot_SN} that $\EdotSN \propto \Sigg \eps$ and that $\Edot_{\rm turb}\propto\Sigg$, so for the fiducial choice of parameters $\Edot_{\rm turb}/\EdotSN \approx (\eps/0.01)^{-1}/3$ and is independent of $\Sigg$.  

Fig. \ref{fig:homog_eta_E_leftover} shows the time evolution of $\etacool$ for the \emph{homogeneous} $\Sigg = 30 \, \mparea$, $\Mcl = 10^{4.5}\, \msun$ simulation. For comparison the energy loading at the top and bottom of the box $\eta_{\rm E}$ is also shown (eq. \ref{eq:eta_def}), as well as $\Ehat$ and $\etacool$ for the matching unstratified simulation. 
For the first $\sim 7$ Myrs $\etacool$ is similar in both the stratified and unstratified simulations, but once the first fingers of bubble material reach the disc's edge and begin to expand freely into the low density medium above and below the disc the amount of energy lost to cooling drop dramatically and the energetics of the stratified and unstratified simulations differ significantly. The energy that is not radiated away in the unstratified simulation goes into expanding and heating the super-bubble, albeit with diminishing efficiency, as reflected by $\Ehat$. However, in the stratified simulations up to 50 per cent of the energy from SNe is not radiated away and is instead carried away by the wind and ends up leaving the domain. This can be seen by the close correspondence of $\etacool$ and $\eta_{\rm E}$. 

\begin{figure}
\includegraphics[width=\columnwidth]{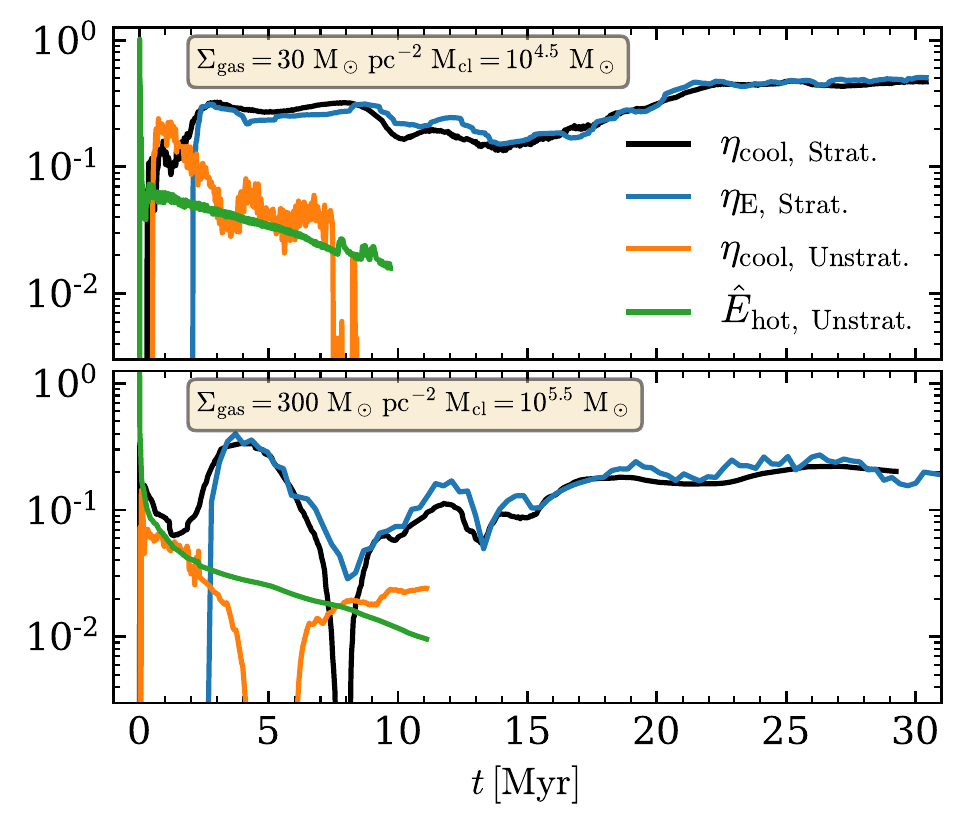}
\caption{Time evolution of $\etacool$ for the stratified (black) and unstratified (orange) turbulent simulations, compared with $\etaE$ leaving the top and bottom of the stratified simulation domain (blue line), and $\Ehat$ from the unstratified simulation (green line). The top panel shows the $\Sigg = 30 \, \mparea$, $\Mcl = 10^{4.5}\, \msun$ simulation and the bottom panel shows the $\Sigg = 300 \, \mparea$, $\Mcl = 10^{5.5}\, \msun$ simulation (both correspond to $\eps = 0.03$).
$\hat{E}_{\rm hot}$ and $\etaE$ trace $\etacool$ because the injected energy not radiated away goes into the energy of the bubble and the wind. The striking divergence of the stratified and unstratified simulations' $\etacool$ after $\sim 3$ Myr when the bubble breaks out demonstrates the efficient venting of SNe energy once a channel through the ISM has been cleared.}
\label{fig:eta_E_leftover}
\end{figure}

Fig. \ref{fig:eta_E_leftover} shows the same quantities as Fig. \ref{fig:homog_eta_E_leftover} but for the \emph{turbulent} $\Sigg = 30 \, \mparea$, $\Mcl = 10^{4.5}\, \msun$ and $\Sigg = 300 \, \mparea$, $\Mcl = 10^{5.5}\, \msun$ simulations. These simulations have $\eps=0.03$, which roughly corresponds to the critical value needed to breakout prior to stalling (see eq. \ref{eq:eps_crit}).
The same finding holds in the turbulent simulations as in the homogeneous simulations: pre-breakout the energetics of the stratified and unstratified simulations are similar while post-breakout they differ dramatically. Relative to the homogeneous simulations the turbulent simulations breakout sooner owing to their lower median densities. Additionally they exhibit more variability in $\etacool$ and $\etaE$ post-breakout. This variability is due to the fact that the massive cold ($T=10^2$ K) clumps in the turbulent simulations -- absent in the homogeneous simulations -- are able to partially, or sometimes completely, re-seal the vent through the ISM that the cluster has carved out. When this occurs the energy released by the cluster is spent on shredding the clump and carving a new vent out of the ISM. This process is inherently sensitive to the properties of the turbulent ISM since that is what sets the flux of cold clumps into the bubble/vent region. We explore in Appendix \ref{app:turb} how our results vary with different turbulent driving realizations. Four otherwise identical simulations with different turbulent realizations yielded a range of $\etaE$ and $\etacool$ on the order of a factor of $\sim$3. The case shown in the top panel of Fig. \ref{fig:eta_E_leftover} lies in the middle of the spread.

\begin{figure}
\includegraphics[width=\columnwidth]{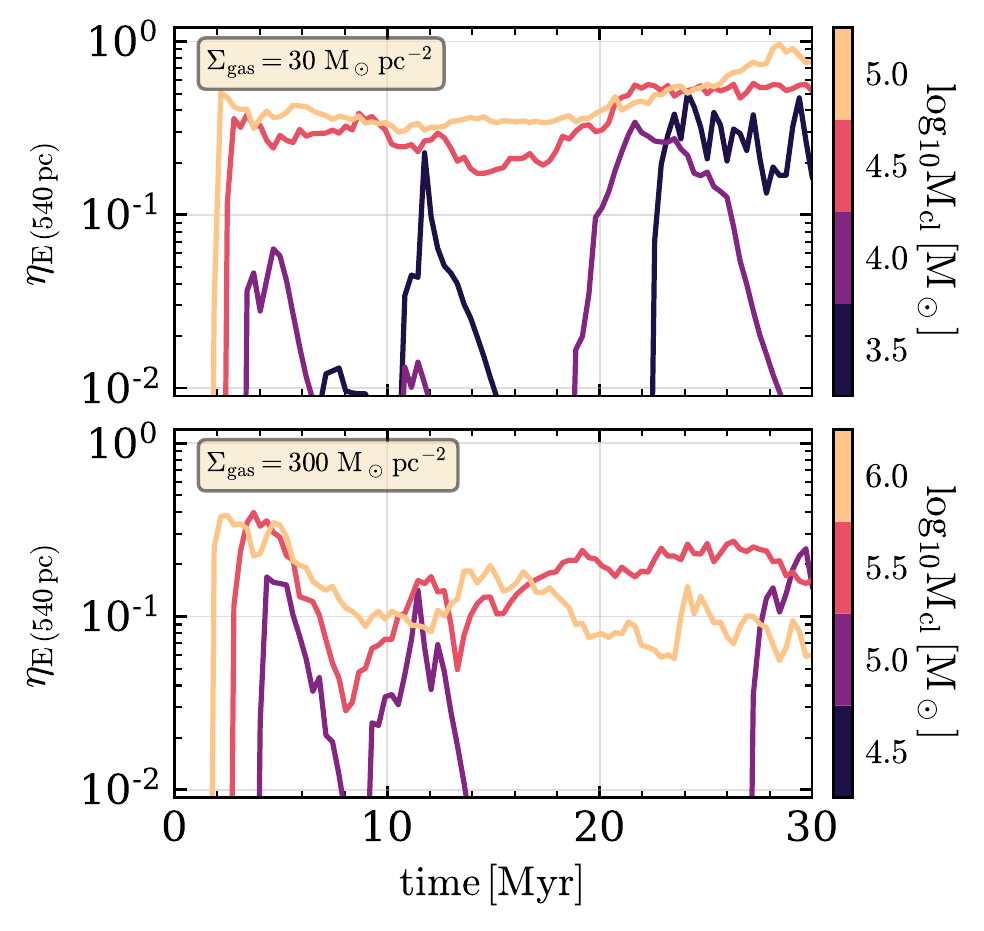}
\caption{Time evolution of the wind energy loading, $\etaE$, measured 540 pc from the disc midplane -- the edge of computational domain -- for $\Sigg = 30$ (top) and $300\ \mparea$ (bottom) simulations. At each surface density clusters with masses corresponding to $\eps = 10^{-2.5},\ 10^{-2},\ 10^{-1.5},$ and $10^{-1}$ are shown. For $\eps \gtrsim 10^{-1.5}$, which corresponds to $\Mcl = 10^{4.5}$ and $10^{5.5}\ \msun$ for $\Sigg = 30$ and $300\ \mparea$ respectively, $\etaE \gtrsim 0.1$ after the initial breakout of the bubble. At lower $\eps$ the bubbles are only able to breakout for short periods of time when the turbulent fluctuations are favorable leading to much lower values of $\etaE$.}
\label{fig:etaE_time}
\end{figure}

\begin{figure}
\includegraphics[width=\columnwidth]{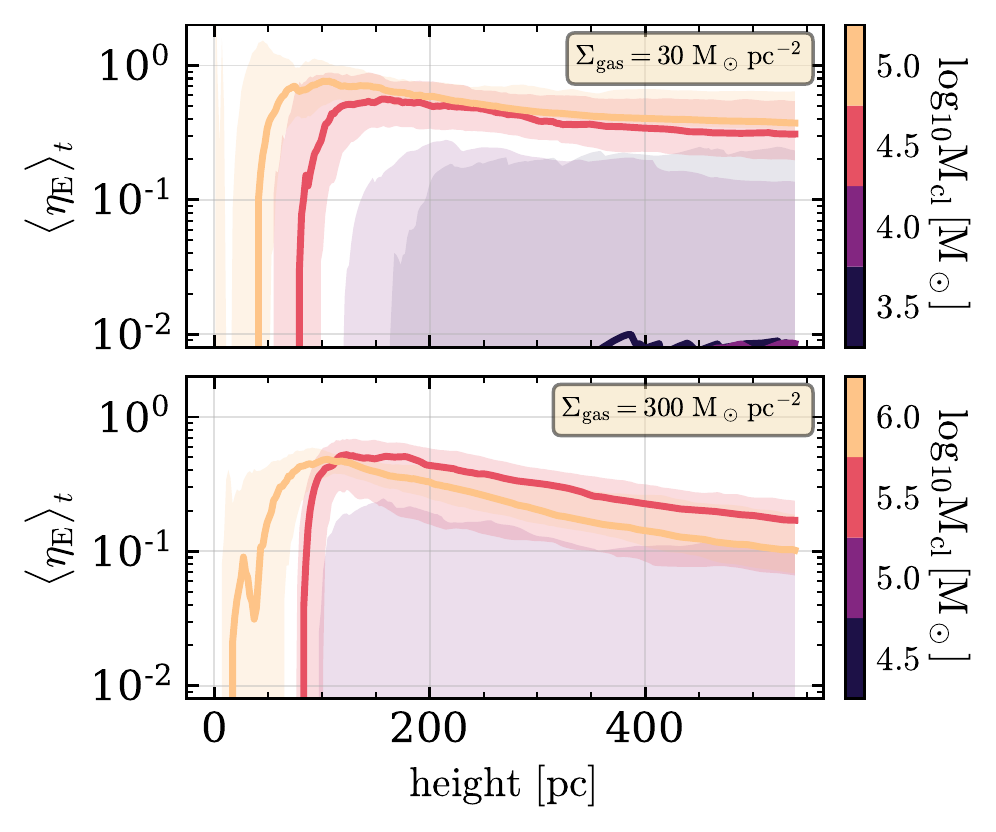}
\caption{Vertical profiles of the time average energy loading $\etaE$ of the $\Sigg = 30$ (top) and $300\ \mparea$ (bottom) simulations. The shaded region denotes the one sigma range of scatter over time. For the sake of clarity we show only the energy loading of outflowing material because within the disc, $|z|<h$, the turbulent motions lead to large variations, and beyond the disc, $|z|>h$ the energy of the outflow is indistinguishable from that of the total. For $\eps \gtrsim 10^{-1.5}$, which corresponds to $\Mcl = 10^{4.5}$ and $10^{5.5}\ \msun$ for $\Sigg = 30$ and $300\ \mparea$ respectively, $\etaE$ is large, $\gtrsim 0.1$, and falls by at most a factor of 3 from $h\approx 100$ pc to the top of the box at $540$ pc, where as for  $\eps \lesssim 10^{-2.0}$ $\etaE$ is small, $\lesssim10^{-2}$ most of the time.}
\label{fig:etaE_height}
\end{figure}

The exact value of $\etaE$ driven by clustered SNe varies somewhat across the range in $\Sigg$ and $\Mcl$ that we explored, but it is generically true that when the bubble is able to breakout (even if only for a short time while the turbulent fluctuations are favorable) the radiative losses are diminished and the winds carry an appreciable faction of the injected energy ($\etaE \gtrsim 0.1$). Fig. \ref{fig:etaE_time} shows the time evolution of $\etaE$ measured at the top and bottom of the domain, $\pm 540$ pc from the disc midplane, for the full range of turbulent stratified simulations. For both surface densities the more massive the cluster the sooner it breaks out of the disc. There are fluctuations in the outflow rate due to changes in the turbulent surroundings. This is most apparent in the lowest $\Mcl$ ($\eps \leq 0.01$) simulations that only break out for short periods when there happens to be a lower ambient density.

Fig. \ref{fig:etaE_height} shows the time averaged vertical $\etaE$ profile for the stratified turbulent simulations. The shaded regions show the one sigma temporal variation in the outflow rates. The lowest mass clusters only occasionally power enough of an outflow for its shaded region to make it into the plotted range and its mean is down around $\etaE\lesssim10^{-2}$. In the winds driven by the more massive clusters the value of $\etaE$ drops by $\lesssim 2$ from the edge of disc at 100 pc to the top of the box at 540 pc. This decrease with height is due to mixing and cooling of the material stripped off entrained clouds. However, there can be artificially enhanced cooling due to the geometry of the numerical setup, as was pointed out by \cite{Martizzi2016} and verified by \cite{Fielding2017}. The periodic boundary conditions in the horizontal directions can limit the expansion and adiabatic cooling of the wind, which in turn can prevent the acceleration of the wind and keep it too hot and slow. That being said, the wind temperatures produced by the clustered SNe are much higher than the wind temperatures produced by the randomly distributed SN studied by \cite{Martizzi2016} and as such cooling is much less efficient.

\begin{figure}
\includegraphics[width=\columnwidth]{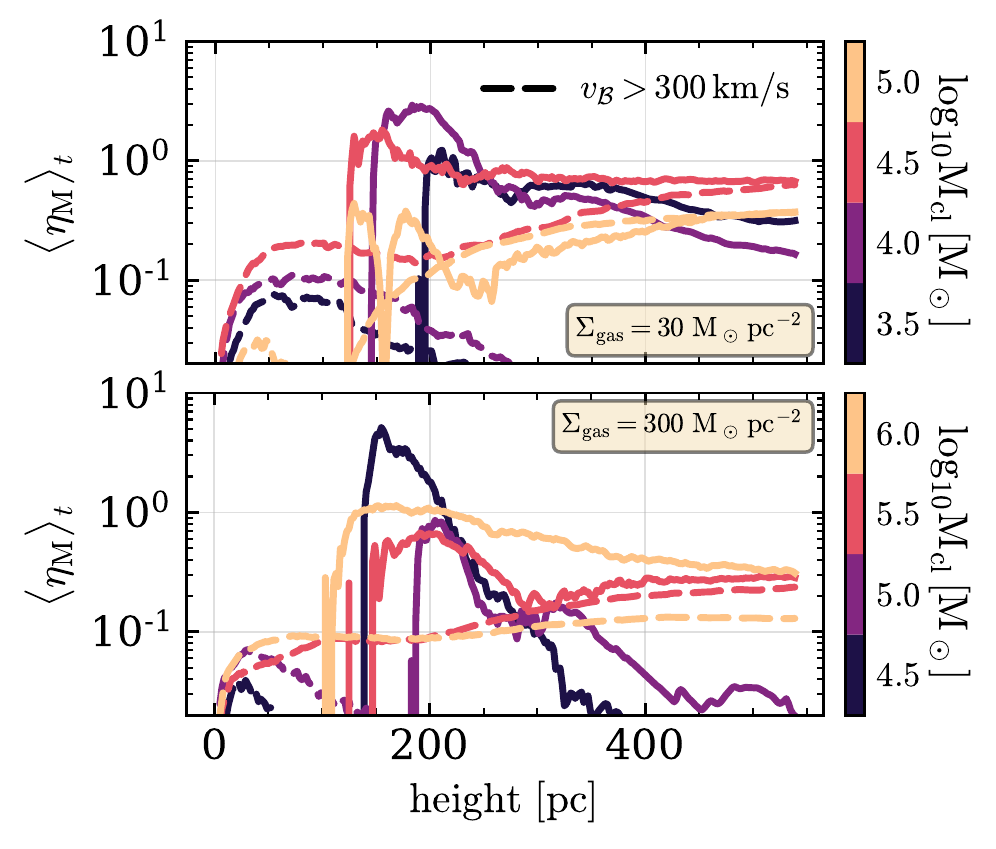}
\caption{Vertical profiles of the time average mass loading $\etaM$ of the $\Sigg = 30$ (top) and $300\ \mparea$ (bottom) simulations. The solid lines show the net (outflowing minus inflowing) mass flux for all of the gas and the dashed lines correspond to only material with $\vB > 300\ \kms$. The total $\etaM$ is not shown within the disc ($|z| < 100$ pc) since the turbulent motions dominate there. The higher $\vB$ material would be able to reach well out into the halo or beyond. In some cases $\etaM$ above the given $\vB$ increases with height due to mixing of low density high energy material with higher density lower energy material as cold clouds are shredded/entrained. }
\label{fig:etaM_height}
\end{figure}

For galaxy formation and the chemical evolution of galaxies it is important to know not just the energy carried by galactic winds but the potentially sizable mass removed from the ISM out into the CGM. 
The solid lines in Fig. \ref{fig:etaM_height} show the time averaged $\etaM$ vertical profiles across the full range of $\Sigg$ and $\Mcl$. The $\etaM$s are not plotted in the midplane ($|z|<100$ pc) to remove confusion caused by the turbulent motions within the disc. Interpreting the $\etaM$ values is further complicated by `fountain' flows where gas is lifted out of the disc but falls back before leaving the domain. 
To account for this, we focus on the highest energy phase of the wind. Specifically we use the fact that the Bernoulli parameter is constant along flow lines (neglecting cooling) to define a `Bernoulli velocity,'
\be \label{eq:vB}
\vB = \sqrt{2\left(\frac{1}{2}v^2 + \frac{\gamma}{\gamma-1}\frac{P}{\rho} \right)}.
\ee
Comparing $\vB$ to the escape velocity gives an estimate for how far the material can go. The dashed lines in fig. \ref{fig:etaM_height} show the portion of $\etaM$ that has $\vB>300\,\kms$, which is the portion of the wind that has the potential to make it far out in the halo and beyond. The high $\vB$ component actually increases with radius. This is due to the mixing of low $\vB$ (mostly cold) material into the high $\vB$ (mostly hot) material. 
In some cases $\etaM$ of the high $\vB$ component is larger than the $\etaM$ of all of the material due to fall back of lower $\vB$ material.
For the higher mass clusters with $\eps \gtrsim 0.03$ ($\Mcl \gtrsim 10^{4.5},\ 10^{5.5}\ \msun$ for $\Sigg = 30,\ 300\ \mparea$, respectively), by a height of 300 to 400 pc the majority of the outflowing material has $\vB>300\,\kms$ and the profiles of total and high $\vB$ components have mostly leveled off. The winds from these higher mass clusters have $\etaM \sim 0.3-1$.

\begin{figure}
\includegraphics[width=\columnwidth]{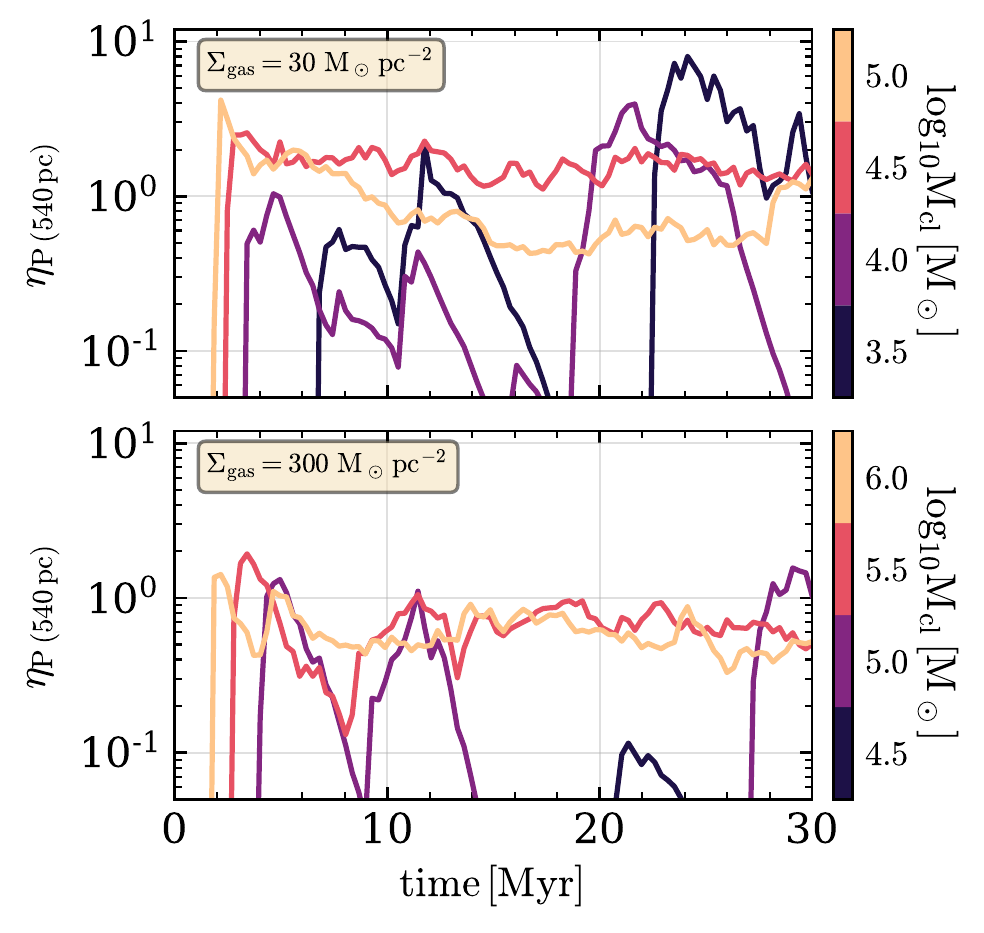}
\caption{Time evolution of the wind momentum loading, $\etaP$, measured 540 pc from the disc midplane -- the edge of computational domain -- for $\Sigg = 30$ (top) and $300\ \mparea$ (bottom) simulations. At each surface density clusters with masses corresponding to $\eps = 0.003,\ 0.01,\ 0.03,$ and $0.1$ are shown. For $\eps \gtrsim 0.03$, which corresponds to $\Mcl = 10^{4.5}$ and $10^{5.5}\ \msun$ for $\Sigg = 30$ and $300\ \mparea$ respectively, $\etaP \sim 1$ after the initial breakout of the bubble. At lower $\eps$ the bubbles are only able to breakout for short periods of time when the turbulent fluctuations are favorable leading to lower average values of $\etaP$.}
\label{fig:etaP_time}
\end{figure}

\begin{figure}
\includegraphics[width=\columnwidth]{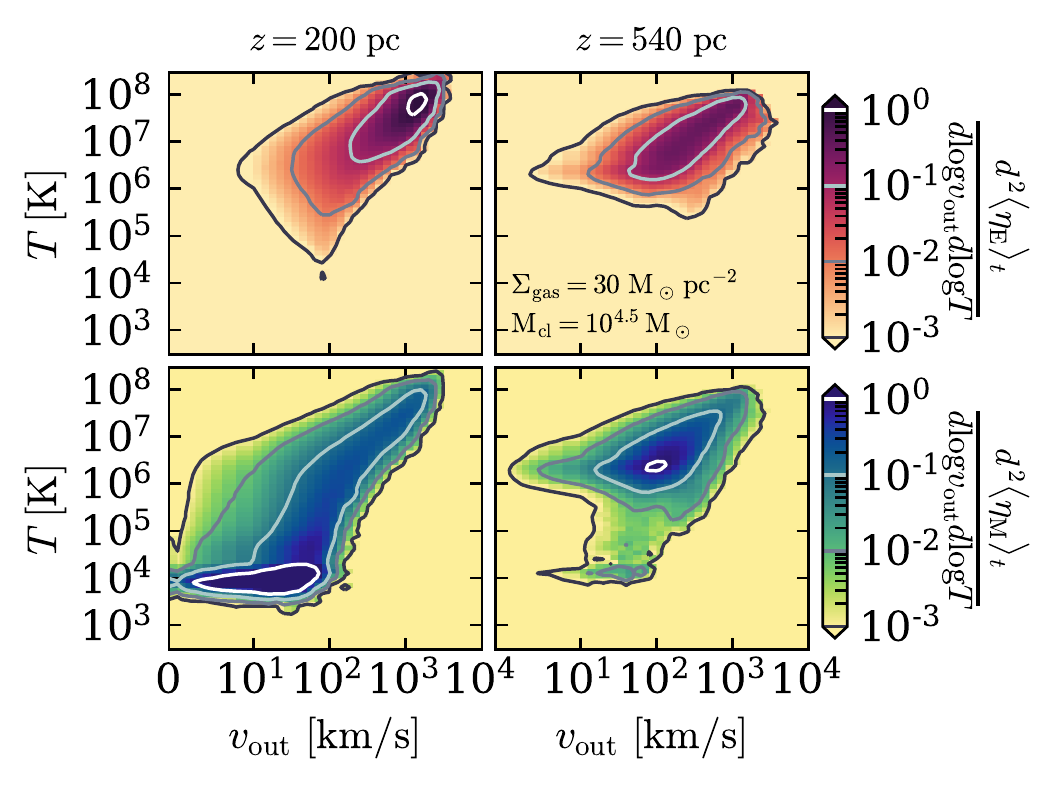}
\includegraphics[width=\columnwidth]{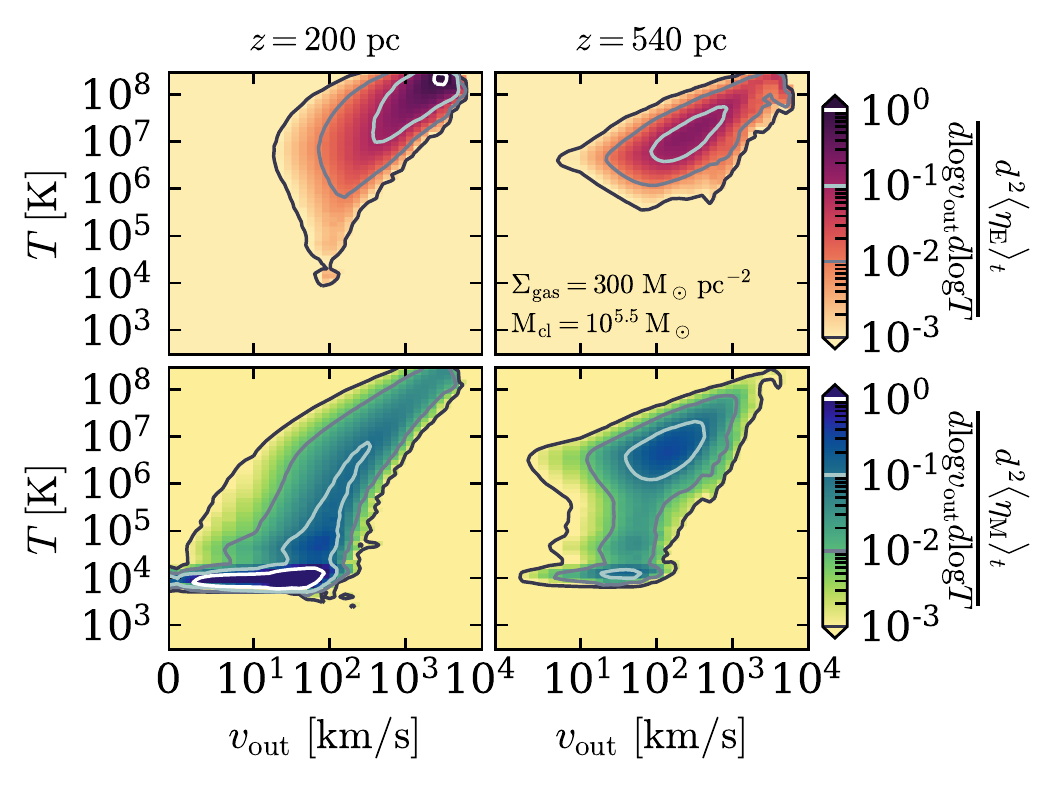}
\caption{Time averaged $\etaE$ (top row) and $\etaM$ (bottom row) per two dimensional logarithmic bin in temperature and outward velocity just above the disc at 200 pc (left column) and at 540 pc, the edge of the computational domain (right column) in the $\Sigg=30\,\mparea$ $\Mcl=10^{4.5}\,\msun$ (top) and $\Sigg=300\ \mparea$ $\Mcl = 10^{5.5}\ \msun$ (bottom) ($\eps=0.03$) simulations. The contour lines are added to guide the eye and are separated by a factor of 10. The energy flux is dominated by the fast hot component at all heights. The mass flux is dominated by the cooler slower phase close to the disc, which mostly drops out or is mixed into the hotter phase so that by the top of the box most of the mass is in the fast hot phase. In the $\Sigg=300\ \mparea$ simulation there is also a notable $T = 10^4$ K component of the wind at 540 pc with $\vout \sim 30 \kms$ that carries a mass flux of $\etaM \sim 0.02$.}
\label{fig:deta_dlog_Sig30}
\end{figure}

In addition to mass and energy the wind carries significant momentum. Analogous to the energy and mass loading we can define a momentum loading 
\be\label{eq:etaP}
\etaP = \frac{\dot{P}_{\rm wind}}{\PdotSN} = \frac{\dot{P}_{\rm wind}}{\Mej\,\vej\,\dot{N}_{\rm SNe}},
\ee
where $\Mej=3\,\msun$ is the mass ejected with each SNe, and $\vej\approx 2.6\times10^3\,\kms$ is the average velocity of the SN ejecta in our simulations. Fig. \ref{fig:etaP_time} show the time evolution of $\etaP$ for all of the turbulent stratified simulations. As with $\etaE$ and $\etaM$, massive clusters with $\eps \gtrsim 0.03$, which are able to breakout, drive winds with significant momentum loading $\etaP \sim 1$, whereas the winds driven by lower mass clusters carry significantly less momentum. There are, however, times when the momentum flux from the lowest $\Mcl$ simulations is high, which occurs intermittently when the conditions in the turbulent ISM are favorable for breakout. 
For comparison, an individual SN remnant in an unstratified ISM effectively has $\etaP \sim 20-40$ due to work done in the Sedov-Taylor phase. 
The fact that post-breakout $\etaP\sim1$ highlights that the energy of the SNe is not going into accelerating ISM material but instead escapes out into the halo. This energy is available to accelerate the inner CGM material and potentially prevent accretion onto the galaxy.

An observationally important question is how much of the wind is hot (harder to observe) or cold (easier to observe). The majority of galactic wind observations are of ionic species that trace cold gas. Likewise, the most readily observable species in the CGM trace cold gas. Understanding how the mass and energy are partitioned between the phases may allow us to better understand the unobserved phases of galactic winds (and the CGM) from observations of a given phase. 
Fig. \ref{fig:deta_dlog_Sig30} demonstrates how the mass and energy fluxes are distributed in temperature and velocity space, showing the amount of $\etaE$ and $\etaM$ per two dimensional logarithmic bins in temperature and outward velocity $\vout$ at two heights for the $\eps=0.03$ turbulent stratified simulations. These distributions are representative of most of the winds in our simulations albeit with minor quantitative variations. Just above the disc (left columns) and at the edge of the box (right columns) the majority of the energy is carried by hot ($T>10^6$ K), fast moving ($v_{\rm out} > 100\,\kms$) gas that has a high enough $\vB$ to escape far out into the halo. Between 200 and 540 pc there is a shift to lower temperature and velocities due to mixing, cooling, and gravity. On the other hand, near the disc (200pc) the mass loading is dominated by much cooler gas ($\sim 10^4$ K) moving outward at $\sim30\ \kms$. By 540 pc this cooler phase of the wind has mostly fallen back or mixed into the hotter phase.
There is, however, a non-zero cold component of the outflow that is moving at nearly the same speed as the hot phase -- more so in the $\Sigg = 300\ \mparea$ simulation than in the $\Sigg = 30\ \mparea$ simulation. This cold component may further out in the halo as the hot medium cools in the wake of the cold clouds \citep[e.g.,][]{Thompson2016, Schneider+18, Gronke+18}.

\section{Discussion and Conclusion} \label{sec:discussion}
\subsection{Summary}

Galactic winds are observed to emanate from a wide range of galaxies and play a critical role in explaining many  global galaxy scaling relations such as the stellar-mass to halo-mass relation and the mass-metallicity relation. The energy injected by SNe into the ISM is one of the most promising mechanisms for driving galactic winds. In this paper we have used numerical simulations, motivated by analytic arguments (see \S \ref{sec:analytic}), to study how spatially and temporally clustered SNe inflate hot super-bubbles in the ISM that can, under a range of conditions, breakout of the disc, vent a large fraction of the injected SNe energy, and drive powerful winds. 

Our numerical simulations targeted gas surface densities of $\Sigg = 30$ and $300\ \mparea$ that are appropriate for vigorously star forming galaxies. At each surface density we studied how changes to the number of SNe in a cluster (parameterized by the cluster mass $\Mcl$ or equivalently the cluster formation efficiency $\eps$; see Table \ref{table:sims}), which sets the time between successive SNe $\Delta t_{\rm SN}$, changed the evolution of the bubble and its ability to breakout. We ran simulations both with and without an external gravitational potential. The simulations without the external potential and the resulting stratification enabled us to isolate the pre-breakout evolution, while the stratified simulations allowed us to study the breakout process and the post-breakout evolution. Similarly, we adopted two choices for the phase structure of the ISM: a homogeneous $10^4$ K ISM, and a more realistic turbulent, multi-phase ISM. The homogeneous simulations help guide physical intuition because of their straightforward interpretation, while the turbulent simulations enabled us to capture the evolution in a more realistic environment including the interaction of the hot super-bubble/wind fluid with cold dense clouds in the ambient ISM. 

There are two possible conditions that determine whether SNe driven bubbles can breakout of a galactic disc. The first is whether the bubble can reach a size of order the disc scale height before reaching pressure equilibrium. The second is whether the bubble can do so prior to the cessation of SNe at $\tsn \sim 30$ Myr. Using the unstratified subset of simulations we confirmed analytic expectations (eq. \ref{eq:Resc} - \ref{eq:vesc}) that the first of these conditions is more stringent (Fig. \ref{fig:r_sh_evo}) and that there is a critical cluster formation efficiency $\eps \sim 0.03$ (or, equivalently, a critical $\Mcl$ or $\Delta t_{\rm SN}$) that determines whether a super-bubble will breakout \citep[see also][]{Kim2017}. 
While the super-bubble is confined within the ISM radiative losses remove between 90 and 99 per cent of the injected energy and leave a hot gas mass of only $\lesssim 10$ per cent of the mass of stars formed (see Figs. \ref{fig:Ehat} and \ref{fig:Mhat}). This efficient cooling seems to argue against the ability of clustered SNe to drive powerful winds \citep{Kim2017}. 

Our stratified simulations, however, uncovered a crucial change in cooling once the super-bubble breaks out and the wind can expand more unimpeded. When the cluster is massive enough for its super-bubble to breakout it blows a `chimney' through the ISM that enables a large fraction ($\sim 0.1-0.6$) of the energy injected by the cluster's SNe to vent into the region above the disc and out into the CGM. During this venting phase cooling is much less effective. This qualitative difference in the energetics of the stratified simulations relative to the unstratified simulations strongly supports clustered SNe as a primary driver of galactic winds (see Figs. \ref{fig:homog_eta_E_leftover} and \ref{fig:eta_E_leftover}). 
The efficient venting is also reflected in the momentum of the wind, which carries roughly the same amount of momentum as is injected by the SNe ($\etaP \sim 1$; see Fig. \ref{fig:etaP_time}). By contrast in unstratified simulations that cannot vent, significant work is done by the SNe and the momentum is boosted by $\sim 30 - 300$ \citep[e.g.,][]{Gentry2017,Kim2017}.  
In addition to the large energy the winds are also significantly mass loaded with $\etaM\sim0.5-1$. Importantly, a large fraction of the mass and energy carried by the wind has the potential to escape far out into the halo (as quantified by having Bernoulli parameters corresponding to speeds $>300\ \kms$; see Fig. \ref{fig:etaM_height}). 

\subsection{Implications and Application to Observations}\label{sec:disc_gal_impl}

Although our simulations predict the energy, momentum, and mass loading of galactic winds ($\etaE, \etaP, \& \, \etaM$ in Figures \ref{fig:etaE_time}-\ref{fig:etaP_time}; see Table \ref{table:definitions}), we believe that the energy and momentum content of the wind are more robust and more useful in diagnosing the importance of winds for galaxy formation.   The primary reason for this is that as the wind propagates into the CGM, $\eta_E$ ($\eta_P$) will be the key conserved quantity if radiative cooling is not (is) important as the wind interacts with the inner regions of the CGM (see, e.g., \citealt{Lochhaas+18}).   By contrast, $\eta_M$ is not a conserved quantity since the wind sweeps up mass as it propagates out into the halo.   In particular, there is sometimes confusion regarding the interpretation of the very large mass loadings $\eta_M \gg 1$ required to explain the low masses of dwarf galaxies in cosmological simulations and semi-analytic models.  To a large extent these large mass-loadings are halo scale quantities, not galaxy scale quantities.  This distinction is related to the distinction between preventive and ejective feedback often discussed in the literature \citep[e.g.,][]{Dave+12}. In low mass galaxies, winds with $\eta_E \sim 1$ and $\eta_M \sim 1$ on galaxy scales (due to efficient venting of SNe like that found here) can prevent accretion of the CGM onto the galaxy, thus explaining the low stellar mass to halo mass of low mass galaxies.  This can effectively correspond to $\eta_M \gg 1$ averaged over the halo.   In fact, X-ray observations of galactic winds rule out $\eta_M \gg 1$ on galaxy scales in actively star forming galaxies \citep{Zhang2014}, strongly supporting a physical picture like that advocated here.   That being said, some consequences of galactic winds for galaxy formation are sensitive to $\eta_M$ on galaxy scales.  This includes, in particular, the chemical evolution of galaxies and the mass-metallicity relation, which depend on the fraction of mass and metals ejected from galaxies.

One key question we do not address in this paper is the fraction of star formation that occurs sufficiently clustered in space and time for SNe to breakout of galactic disks and drive powerful winds. 
Convolving this fraction with our results on wind strength as a function of cluster mass $M_{\rm cl}$ (or, equivalently, $\epsilon_*$) would determine the overall wind strength from a given galaxy. It is worth stressing that the clustering required to enhance the strength of galactic winds does {\em not} imply that the star formation must occur in bound clusters.   All SNe that are correlated in time on timescales $\lesssim t_{\rm SN} \sim 30$ Myrs and space on lengthscales $\lesssim$ the disc scale-height can overlap, thus enhancing the efficacy of wind driving.    

We now consider the application of our results to the prototypical local starburst M82 and $z \sim 2$ star forming disc galaxies.    M82 has  $\Sigma_g \simeq 3000 \, \mparea$, $t_d \sim 2\times10^6$ yrs, and $\delta v \sim 20 \kms$ \citep{1998ApJ...498..541K,Greco2012}.   Equations \ref{eq:Resc} and \ref{eq:vesc} thus imply $R_s(t = \tsn) \simeq 5 h$ and $v_s(R_s = h) \sim 20 \kms$, such that breakout at $t \ll \tsn$ is plausible, but only for our fiducial $\eps = 0.03 - 0.1$.   The latter corresponds to $\Mcl \sim 0.5-1.5\times10^{6} \,\msun$ for our assumed M82 conditions, consistent with the star cluster masses in M82 \citep{McCrady2007}.   Thus our model argues that clusters like those observed can indeed account for the large energy in the hot wind in M82 inferred from {\em Chandra} observations by \citet{Strickland2009}.   For $z \sim 2$ star forming discs, $\Sigma_g \sim 10^{2-3} \, \mparea$, $\delta v \sim 30-50 \kms$, and $t_d \sim 3 \times 10^7$ yrs \citep{Tacconi2013}.   Equations \ref{eq:Resc} and \ref{eq:vesc} thus imply $R_s(t = \tsn) \simeq 2 h$ and $v_s(R_s = h) \sim 100 \, \kms \gg \delta v$; thus breakout is again likely satisfied, leading to efficient venting of late-time SNe associated with star clusters.  These comparisons support a key role for clustered SNe in driving powerful galactic winds in a wide range of star forming galaxies.

\subsubsection{A Minimum Star Formation Rate Surface Density for Galactic Winds}

To further expand on the implications of these results, we suggest here that the role of clustered SNe in driving galactic winds may set a minimum star formation rate surface density $\dot \Sigma_*$ for galactic winds.

Theoretical models of GMC disruption by star clusters find that  GMCs are harder to disrupt in higher surface density environments \citep{Murray2010}.   This suggest that $\eps$ will be a increasing function of increasing surface density $\Sigma_g$.   For concreteness, consider $\eps = \epsilon_0 \Sigma_g/\Sigma_{\rm max}$ for $\Sigma_g \lesssim \Sigma_{\rm max}$, with $\epsilon_0 \sim 1$ and $\Sigma_{\rm max} \equiv 3000 \, \Sigma_{\rm 3000} \mparea$ \citep{Grudic2017}; the exact functional form assumed here is not critical for what follows.    Using the analytic scalings from \S \ref{sec:analytic}, we then find that breakout $v_s(R_s = h) \gtrsim \delta v$ only occurs if
\be
\Sigma_g \gg \Sigma_{\rm crit,1} \simeq 40 \, f_V \epsilon_0^{-1} \, \tdnorm^{-1} \, \delta v_{10} \, \Sigma_{\rm  3000} \, \mparea
\label{eq:Sigma-thr1}
\ee
It is also useful to rewrite equation \ref{eq:Mcl-min} using equation \ref{eq:mcl}, \ref{eq:n}, and $\eps(\Sigma_g)$, which yields
\be
\Sigg \gg \Sigma_{\rm crit,2} \simeq \, 20 \, f_V^{1/4} \left(\frac{\Sigma_{\rm 3000}}{\epsilon_0 \, \delta v_{10}}\right)^{0.6} \, \tdnorm^{-1.5}  \, \mparea
\label{eq:Sigma-thr2}
\ee
Equations \ref{eq:Sigma-thr1} and \ref{eq:Sigma-thr2} show that i) the SNe associated with star clusters only coherently drive bubbles and ii) the resulting bubbles only breakout out of the galactic disc if the gas surface density of the disc is sufficently large.   
The surface density thresholds in equations \ref{eq:Sigma-thr1} and \ref{eq:Sigma-thr2} 
correspond, via the Kenicutt-Schmidt relation, to a condition on the star formation rate per unit area of the disc required to drive a strong galaxy-scale wind, roughly $\dot \Sigma_* \gg 0.03 \, \sfrarea$.   This is comparable to the observational threshold described by \citet{Heckman2002}.     We predict that it is a correlation between star cluster properties and gas surface density that  ultimately produces a star formation surface density threshold for galactic winds.

\subsection{Comparison to Related Work}
We now discuss our findings in the context of related numerical work, restricting our discussion only to the most similar work. First, we compare to other simulations of clustered SNe in unstratified media, followed by a comparison to simulations of winds launched by (not necessarily clustered) SNe in a stratified medium. 

The three dimensional inhomogeneous unstratified simulations of \cite{Kim2017} are the most directly comparable to our unstratified turbulent simulations. \cite{Kim2017} focused on somewhat lower mean ISM densities, ranging from $n=0.1$ to $10\,\cc$, than we have. When comparing our most similar simulations the $\Mhat$ and $\Ehat$ in our simulations are roughly $\sim 3$ and 10 times higher, respectively, than in theirs. Reassuringly the bubble radii in both of our simulations grow as $t^{1/2}$ appropriate for the momentum-driven regime and have similar normalizations. Nevertheless, it is worthwhile to consider possible explanations for why the values of $\Mhat$ and $\Ehat$ differ. Although our simulations are similar there are differences in the details of how our cooling, photoelectric heating, and SN injection are implemented; and although in both cases the ISM is inhomogeneous in our simulations it is turbulent while theirs is static. Moreover, from a purely computational fluid dynamics standpoint the differences could be due to differences in choice of Riemann solver, reconstruction, or integration scheme (we used the HLLC Riemann solver, with plm reconstruction and a Van Leer integrator), all of which can change the properties of cooling and mixing \citep[e.g.,][]{Martizzi+18, Gronnow+18}.  
All told it is not that surprising that the quantities most sensitive to mixing differ depending on simulation details.

Separately, both 1D and 3D unstratified homogeneous ISM simulations have provided valuable insight into the numerical challenges in obtaining converged results for SNe-driven super-bubble evolution \citep{Yadav+17,Gentry2017,Gentry+18}. The root of this challenge can be traced to how thin the forward shock is once it has cooled. \cite{Gentry2017} showed that the radial momentum per SNe is well converged when using a 1D moving mesh code that can resolve the thin shell with many cells, but when the grid was fixed, even with sub-pc resolution the radial momentum per SNe was an order of magnitude smaller, and not converged with resolution (see their fig. 15). This striking difference may, however, be artificial due to mixing processes not captured in 1D simulations. This includes the Rayleigh-Taylor instability, which we find is important even for a homogeneous ambient medium (see Appendix \ref{app:RT}). 

The 3D homogeneous ISM simulations presented by \cite{Yadav+17} and \cite{Gentry+18} also stress the difficulty in obtaining converged results. Although these simulations are able to capture physical multi-dimensional mixing, these authors concerns about convergence may not be relevant since it only appears at late times $t\sim\tsn$ when the super-bubble would have already broken out of the galactic disc. Moreover, the real ISM is highly inhomogeneous and the mixing is dominated by the cold clumps that are enveloped by/penetrate the expanding super-bubble.  In Appendix \ref{app:res} we look in detail at the super-bubble properties as a function of resolution for both the homogeneous and inhomogeneous simulations over the few Myr timescale before a bubble would break out of the galactic disc. We find the results to be well converged, with the inhomogeneous simulations showing more of a resolution dependence, as well as enhanced mixing and cooling relative to the homogeneous simulations. 

In addition to the work on clustered SNe in an unstratified medium, much work has gone into simulating the winds launched by SNe in a stratified medium. \cite{Girichidis+16} performed a related study, measuring the difference between detonating SNe at density peaks, randomly distributed in the ISM, or clustered. They found significantly higher $\etaM$ than we have, but there are numerous differences in our methods that can account for this. Notably, their disc, which had $\Sigg=10\ \mparea$ was thinner with a gas scale height of $30$ pc and as shown in eq. \ref{eq:vesc} thinner discs are easier to breakout of. Moreover the ISM turbulence required to support this scale height was not driven  externally but instead generated by the SNe themselves. Without initial turbulence the disc initially becomes even thinner. \cite{Fielding2017} also studied how clustering SNe increases their ability to drive powerful winds. By systematically increasing the degree of clustering they showed that in cases where randomly distributed SNe launch effectively no wind at all clustered SNe can drive powerful ($\etaE\sim0.5$ and $\etaM\sim$ few) winds. The winds in \cite{Kim+18}'s more physically realistic simulations of SNe in a stratified medium of are comparable to what we find here, albeit a factor of a few lower in $\etaE$ and $\etaM$. Their star formation and subsequent SN locations are handled self-consistently, so clustering can arise naturally. However, their simulations probe lower surface density discs with $\Sigg=10\ \mparea$.

\subsection{Missing Physics}\label{sec:missing_phys}

There are numerous complex physical processes at play in the ISM that together determine the properties of its different phases and contribute to setting the properties of galactic winds. In keeping with the idealized nature of our numerical simulations we have limited ourselves to a restricted set of these processes. This enabled us to keep the problem tractable and the interpretation of the results relatively straightforward. There are, however, several processes that we have not considered here that may have an important impact on the galactic winds driven by clustered SNe, in particular, magnetic fields, thermal conduction, self-gravity, and additional feedback processes. 

The inclusion of magnetic fields may change the winds driven by clustered SNe by changing the pre-breakout evolution and the shear-flow mixing. Within the Milky Way's diffuse ISM magnetic fields are observed to be in rough equipartition with the thermal pressure. This large additional energy density in the ISM can change the early evolution of a bubble while it is still confined within the disc. \cite{Gentry+18} demonstrated that magnetic tension forces can drain momentum from an expanding bubble. Moreover, magnetic fields can suppress mixing by stabilizing shear instabilities. During the pre-breakout phase this is relevant when cold clouds are enveloped by the hot bubble and during the development of the Rayleigh-Taylor instability \citep{Stone+07}. Magnetic fields are also likely important with regards to the shredding and entraining of cold clouds by the wind post-breakout. As seen in Fig. \ref{fig:deta_dlog_Sig30} only a small fraction of the mass carried by the wind is at low temperature, but many observations of galactic winds show a sizable cold component. Magnetic fields can dramatically prolong the lifetime of a cold cloud in a hot wind \citep{McCourt+15}. Furthermore, suppressed mixing in the wind may reduce the degree of radiative losses the wind suffers. 

The problem at hand inherently has many regions with steep temperature gradients, which would lead to large conductive fluxes. Conduction could impact both the phase structure of the ISM and therefore the expansion of the super-bubble as well as the mixing of the cold clouds as they are entrained and shredded. 

Including self-gravity would cause the dense structures in the ISM to be more tightly bound and may impact their survival when interacting with the bubble/wind material. Moreover, by including self-gravity, star formation would be tied to gravitational collapse, thereby giving a self-consistent relation between a clusters location and the ISM density and velocity field. This may be important because the proximity of the cluster to massive cold clouds can have a sizable impact on the evolution of the bubble and strength of the wind (see Appendix \ref{app:turb}). 

In recent years much work has gone into understanding the role of cosmic rays in launching galactic winds. Cosmic rays introduce an appreciable pressure gradient which can lift material above the disc where it can be more easily unbound by SNe \citep[e.g.,][]{Salem+14,Girichidis+16a,Butsky+18}. Therefore the combined effect of SNe and cosmic rays may further increase the strength of galactic winds. Likewise, other feedback processes, such as photoionization, radiation pressure, and stellar winds, might clear gas out around star clusters prior to the onset of SNe. This would effectively increase $\eps$ by decreasing $\Sigg$ around a given cluster, making breakout and strong winds more likely.

As stressed in this section, there are many effects that may play a role in determining the detailed properties of galactic winds.  However, there is no reason to think that the processes we have omitted from this current study would qualitatively change our primary finding that post-breakout cooling is significantly reduced and clustered SNe can drive powerful winds by efficiently venting a large fraction of their energy out of the ISM. That said, the quantitative details of the wind properties are likely subject to change and our results should be considered instructive guides rather than the final word on the subject.

\subsection{Acknowledgments}
DF and EQ would like to thank Chris J. White, Chang-Goo Kim, Chris McKee, Evan Schneider, Eve Ostriker, Todd Thompson, and Claude-Andr\'e Faucher-Gigu\`ere for useful conversations regarding the physical processes relevant to launch galactic winds and their numerical implementations. 

We thank the Simons Foundation and the organizers of the workshop \emph{Galactic Winds: Beyond Phenomenology} (J. Kollmeier and A. Benson) where this work germinated.

EQ was supported in part by a Simons Investigator Award from the Simons Foundation and by NSF grant AST-1715070. DM was supported in part by the Swiss National Science Foundation postdoctoral fellowship grant P300P2\_161062, in part by NASA ATP grant 12-APT12-0183 and in part by the CTA and DARK-Carlsberg Foundation Fellowship.

This work used the Extreme Science and Engineering Discovery Environment (XSEDE) comet at SDSC through allocation TG-AST160020, as well as the savio computational cluster resource provided by the Berkeley Research Computing program at the University of California, Berkeley (supported by the UC Berkeley Chancellor, Vice Chancellor for Research, and Chief Information Officer).

\bibliography{ref}


\appendix
\section{Spatial Resolution Convergence}\label{app:res}

\begin{figure}
\includegraphics[width=\columnwidth]{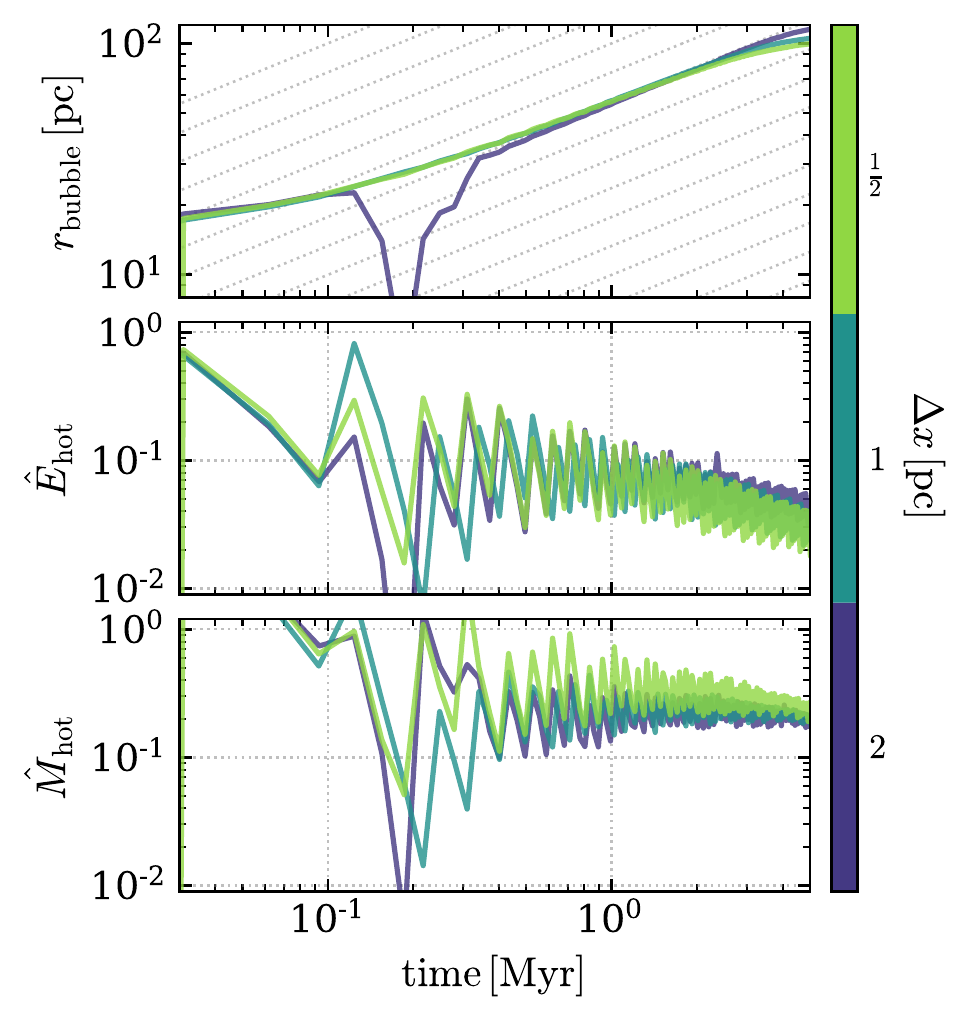}
\caption{Resolution dependence of the evolution of $r_{\rm bubble}$ (top), $\Ehat$ (middle), and $\Mhat$ (bottom) in the turbulent unstratified $\Sigg=30\,\mparea$ $\Mcl=10^{4.5}\,\msun$ simulations. Throughout the $5$ Myr duration of this test the quantities agree extremely well and are (at worst) within a few tens of per cent of each other. The highest resolution simulation has slightly lower $r_{\rm bubble}$ and $\Ehat$ and higher $\Mhat$ as a result of slightly enhanced mixing and subsequent cooling.}
\label{fig:unsrat_inhomog_resolution}
\end{figure}

\begin{figure}
\includegraphics[width=\columnwidth]{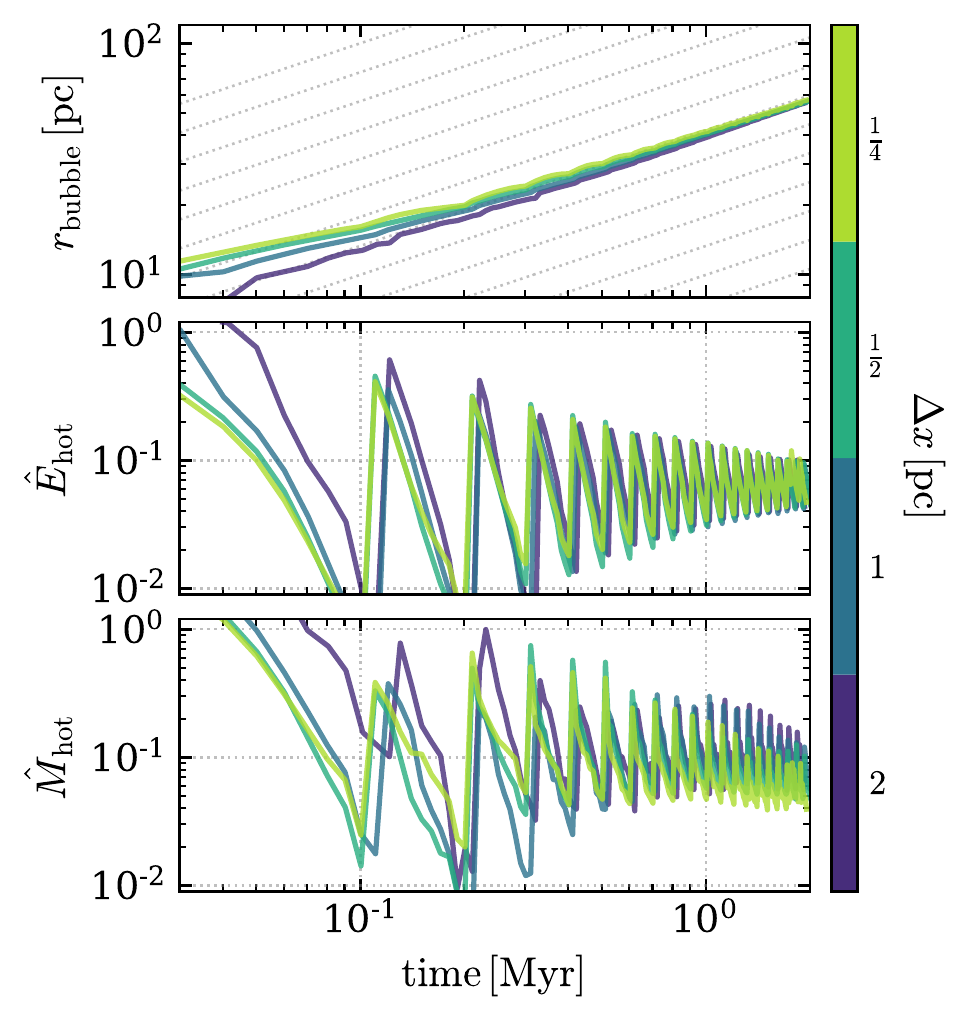}
\caption{The evolution of $r_{\rm bubble}$ (top), $\Ehat$ (middle), and $\Mhat$ (bottom) in homogeneous unstratified $\Sigg=30\,\mparea$ $\Mcl=10^{4.5}\,\msun$ simulations with spatial resolutions ranging from $\Delta x = 0.25$ to 2 pc. These simulations were run in a box 128 pc on a side, so they were stopped after 2 Myr once the bubble had reached the edge of the domain. The agreement of these quantities is excellent across a factor of 8 in resolution.}
\label{fig:unsrat_homog_resolution}
\end{figure}

Previous studies have reached conflicting conclusions regarding the numerical convergence of bubble evolution. This is most likely due to differences in the convergence of 1D versus 3D simulations and differences in numerical techniques. Groups that investigated the evolution of cluster-driven super-bubbles in a homogeneous ISM found that as they decreased the cell size the amount of energy lost to cooling decreased \citep{Yadav+17, Gentry2017}. Alternatively, \cite{Kim2017} simulated the evolution of cluster-driven super-bubbles in an inhomogeneous ISM and found their results to be well converged with spatial resolution. In Appendices \ref{app:unstrat_turb_res} and \ref{app:homog} we investigate the resolution dependence of our results for the unstratified turbulent and homogenous simulations, respectively. In both cases we find our results to be very well converged and discuss briefly why previous homogeneous ISM simulations may have over estimated the resolution dependence. 

In Appendix \ref{app:strat_turb_res} we look at the convergence of the turbulent stratified simulations, focusing on the post-breakout energetics and wind properties. The results at different resolutions agree well -- on the tens of per cent level. 

\subsection{Turbulent Unstratified Simulations}\label{app:unstrat_turb_res}

To test the spatial resolution sensitivity of our turbulent unstratified simulations we ran otherwise identical $\Sigg=30\,\mparea$ $\Mcl=10^{4.5}\,\msun$ simulations with spatial resolutions of $\Delta x = 0.5$, 1 and 2 pc. The initial conditions of the higher resolution simulations were generated by refining the lowest resolution simulation. This guaranteed that any variations we see are not caused by different ISM properties. That being said the subsequent turbulent driving did not use the same random numbers so minor differences may arise at late times because of this. 

Fig. \ref{fig:unsrat_inhomog_resolution} shows $r_{\rm bubble}$ (top), $\Ehat$ (middle), and $\Mhat$ (bottom) for these simulations. The agreement is excellent. The highest resolution simulation has marginally higher $\Mhat$ and lower $r_{\rm bubble}$ and $\Ehat$, at the level of ten per cent or less. This indicates that there is slightly more mixing at higher resolution. Previous studies with inhomogeneous ISMs found similar convergence \citep{Kim2017}.

\subsection{Homogeneous Unstratified Simulations}\label{app:homog}

Our unstratified homogeneous simulations enable us to compare with turbulent simulations to understand what is mediating mixing between the ISM and the super-bubble. 
Previous work that adopted a homogeneous ISM found that their results were not converged \citep{Yadav+17, Gentry2017}, which has raised questions about the robustness of simulations with an inhomogeneous ISM. 

Fig. \ref{fig:unsrat_homog_resolution} shows $r_{\rm bubble}$ (top), $\Ehat$ (middle), and $\Mhat$ (bottom) for $\Sigg=30\,\mparea$ $\Mcl=10^{4.5}\,\msun$ simulations with $\Delta x$ ranging from 0.25 to 2 pc. These simulations were run in a 128 pc box and were stopped prior to the bubble reaching the boundaries at 2 Myr. All three quantities agree exceptionally well. At higher resolution the bubble evolves somewhat faster for the first few tenths of a Myr. However, across a factor of 8 in resolution these quantities vary by at most a few per cent. Therefore even 2 pc resolution is sufficient to capture the initial expansion of the bubble.

\cite{Yadav+17} ran similar three-dimensional homogeneous unstratified simulations and found that at higher resolution the cooling decreased. They, however, focused at late times $t\sim\tsn$ well after the super-bubble would have broken out of a galactic disc. Their Fig. 18 also shows clearly that the resolution dependence is decreasing with increasing resolution in their three-dimensional simulations, which indicates they are approaching convergence by $\Delta x = 1$ pc. Moreover they adopt a ten times larger cluster radius of 100 pc, so it takes much longer for the SNRs to overlap. 
We thus believe our convergence results are reasonably consistent.

However, one dimensional homogeneous unstratified ISM simulations are inconsistent with our results \citep{Yadav+17, Gentry2017}. As we show in the next section this is due in large part to their inability to capture multi-dimensional instabilities that arise in the contact discontinuity separating the bubble and ISM.

\begin{figure*}
\includegraphics[width=6.5in]{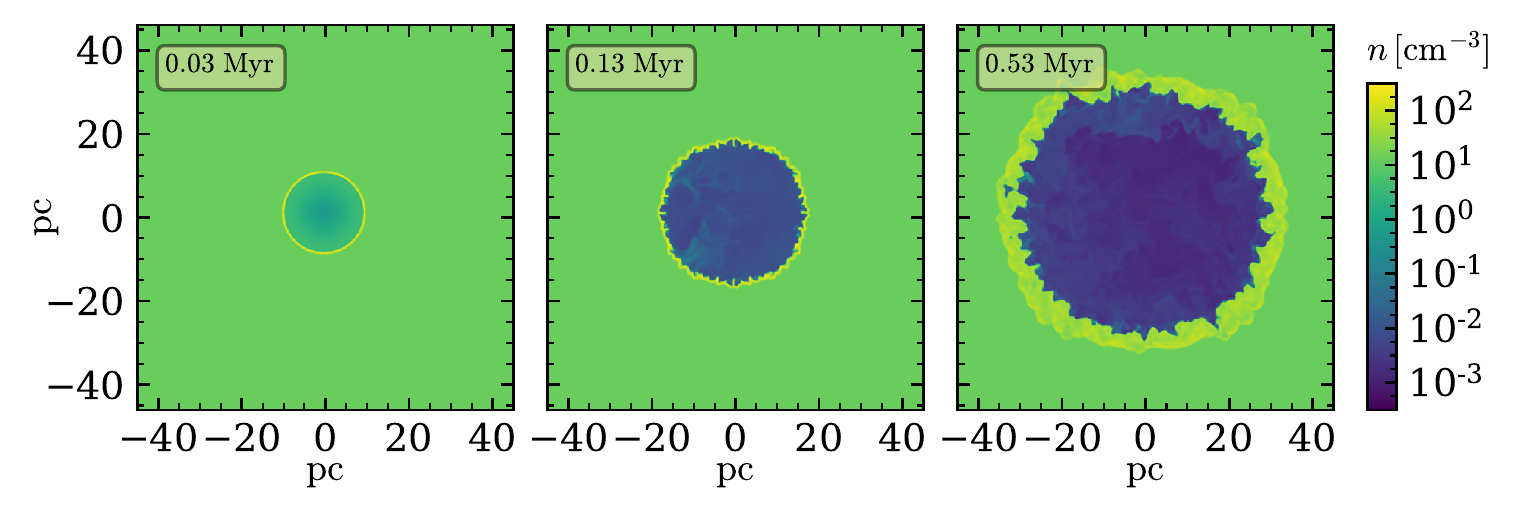}
\caption{Slices showing the number density after the first, second, and sixth SNe in a homogeneous unstratified $\Sigg=30\,\mparea$ $\Mcl=10^{4.5}\,\msun$ simulation with $\Delta x = 0.25$ pc. After the first SN the contact discontinuity separating the bubble and ISM is nearly perfectly spherical. However, after the second SN and all subsequent SNe the contact discontinuity is no longer symmetric but instead is significantly corrugated. The disruption of the contact discontinuity develops due to Rayleigh-Taylor instability.This generates physical mixing between the shocked ISM and the SNe ejecta, and makes the simulations converge much better than the analogous one dimensional simulations.\label{fig:RT}}
\end{figure*}

\subsubsection{Rayleigh-Taylor}\label{app:RT}

The one dimensional homogeneous unstratified ISM simulations performed by \cite{Yadav+17} and \cite{Gentry2017} showed clearly that as the resolution was improved the amount of energy lost to cooling decreased. One dimensional simulations are able to achieve far higher resolution than three dimensional simulations, but they are unable to capture multi-dimensional instabilities. In particular, in this case, the ability capture the Rayleigh Taylor (RT) instability is crucial to accurately model the mixing of the bubble and ISM. As the bubble expands it sweeps up material and forms a thin dense shell. \cite{Weaver1977} demonstrated that in the limit of constant energy and mass injection the contact discontinuity separating the shell and the bubble is stable to RT. However, because the supernovae inject energy sporadically not continuously the shell experiences impulsive pushes after each explosion. These explosions accelerate the less dense bubble material into the more dense shell material -- the density gradient and the pressure gradient have opposite signs -- setting up the conditions for the RT. 

Clear signs of development of RT can be seen in Fig. \ref{fig:RT}, which shows density slices immediately after the first, second and sixth SNe in the 0.25 pc resolution homogeneous unstratified simulation. After the first SN the shell is nearly perfectly spherical and is stable to RT because the shell is decelerating. On the other hand, after all subsequent SNe there are clear signs of disruption to the contact discontinuity due to the growth of RT. The disruption of the contact discontinuity and subsequent mixing is a real physical effect that is not captured in the one-dimensional simulations, which are therefore prone to underestimating the mixing and cooling. This was also confirmed by \cite{Gentry+18}. Although in the simulations shown in Fig. \ref{fig:RT} grid scale noise seeds the growth of the instability, in the turbulent simulations and in the real universe inhomogeneities in ISM are unavoidable and the instability will have ample perturbations to amplify.

\subsection{Turbulent Stratified Simulations}\label{app:strat_turb_res}

\begin{figure}
\includegraphics[width=\columnwidth]{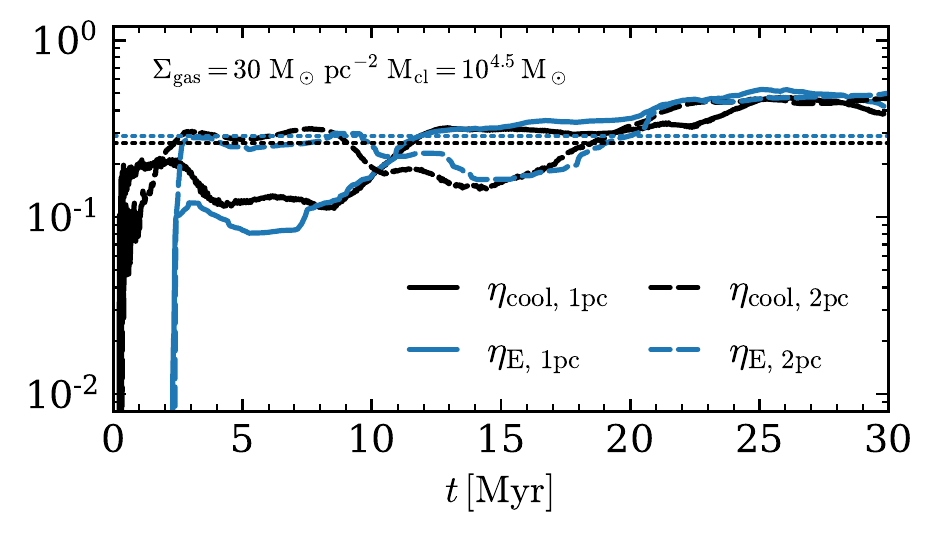}
\caption{The time evolution of $\etacool$ (black) and $\etaE$ (blue) measured at 540 pc for $\Sigg=30\,\mparea$ $\Mcl=10^{4.5}\,\msun$ turbulent stratified simulations with $\Delta x = 1$ pc (solid lines) and 2 pc (dashed lines). Initially the 1 pc simulation has $\sim 2$ more cooling, but during the middle of the simulations this is reversed, and by the end the two have nearly identical $\etaE$ and $\etacool$. The thin dotted lines show the time average $\etaE$. The close agreement indicates that the post-breakout evolution of our simulations is reasonably well converged. \label{fig:etacool_res}}
\end{figure}

We now shift our attention to the numerical convergence of the stratified turbulent simulations. In the previous section we assessed the convergence of the pre-breakout dynamics and energetics and showed that 2 pc resolution is sufficient to accurately capture the bubble evolution. Here instead we focus on the post-breakout evolution convergence, looking at the cooling loses, energy and mass loading, and the phase distribution of the wind. Capturing the shredding and entrainment of cold clumps by a hot wind is well known to be fraught with numerical difficulties, and cloud crushing simulations have shown that with higher resolution cold clouds are shredded more quickly \citep[e.g.,][]{Schneider+17}. Much of the mass-loading of the winds launched in our simulations comes from the shredding of cold clouds, so this resolution dependence could potentially impact our findings. Moreover, changes to the mass loading can potentially change the degree of radiative cooling in the wind. 

To test the dependence of post-breakout dynamics on resolution we re-simulated the $\Sigg=30\,\mparea$ $\Mcl=10^{4.5}\,\msun$ turbulent stratified simulation with twice the spatial resolution, pushing $\Delta x$ down to 1 pc. The initial conditions for the higher resolution simulation were generated by refining the initial conditions of the fiducial resolution simulation. Although the subsequent driving is different, the matched initial conditions ensures that the ISM structures and dynamics are similar. 

Fig. \ref{fig:etacool_res} shows $\etacool$ and $\etaE$ for the $\Delta x = 1$ and 2 pc stratified turbulent simulations. The 1 pc simulation initially cools more and drives a weaker wind by a factor of $\sim 2$, but after about 10 Myr this trend flips and the 1 pc simulation cools less than the 2 pc simulation. For the final 10 Myr of the simulations the two resolutions cool at essentially the same rate and drive comparable winds. The time averaged $\etaE$ and $\etacool$ between these two simulations agree well -- with $\etaE\approx\etacool = 0.26$ and 0.29 for the 1pc and 2pc simulations, respectively (shown with the dotted lines). Given the degree of numerical complications inherent to this problem this the level of agreement is encouraging and supports our primary finding that the energetics change dramatically post-breakout. 

\begin{figure}
\includegraphics[width=\columnwidth]{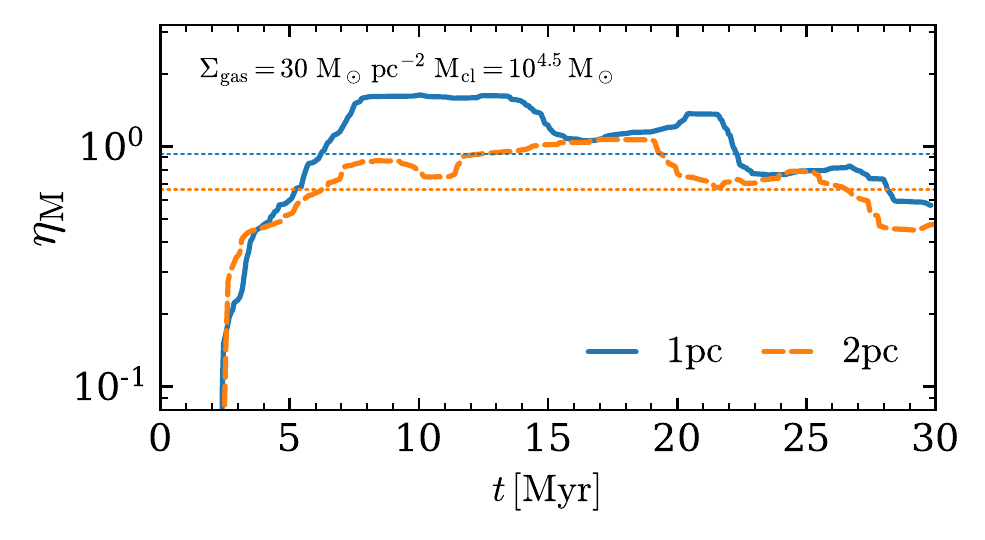}
\caption{The time evolution of $\etaM$ measured at 540 pc for the $\Sigg=30\,\mparea$ $\Mcl=10^{4.5}\,\msun$ turbulent stratified simulation with $\Delta x = 1$ pc (blue solid line) and 2 pc (orange dashed line). The dotted lines show the time average. The mass loading of the higher resolution simulation is systematically higher than the lower resolution simulation by a few tens of per cent. This is likely due to enhanced entrainment of material shredded off of cold clumps that is better captured with higher resolution. Nevertheless the differences are relatively minor and our conclusions are qualitatively unchanged.\label{fig:etaM_res}}
\end{figure}

The mass-loading of the winds, on the other hand, shows a slightly larger dependence on the resolution. Fig. \ref{fig:etaM_res} shows the time evolution of $\etaM$ measured at 540 pc for the two resolutions. Cold clumps in the higher resolution simulation are shredded more efficiently which enhances the mass flux out of the domain. The dotted lines show the time averaged value of $\etaM$, which drops from 0.9 to 0.7 when going from 1 pc to 2 pc resolution.

\begin{figure}
\includegraphics[width=\columnwidth]{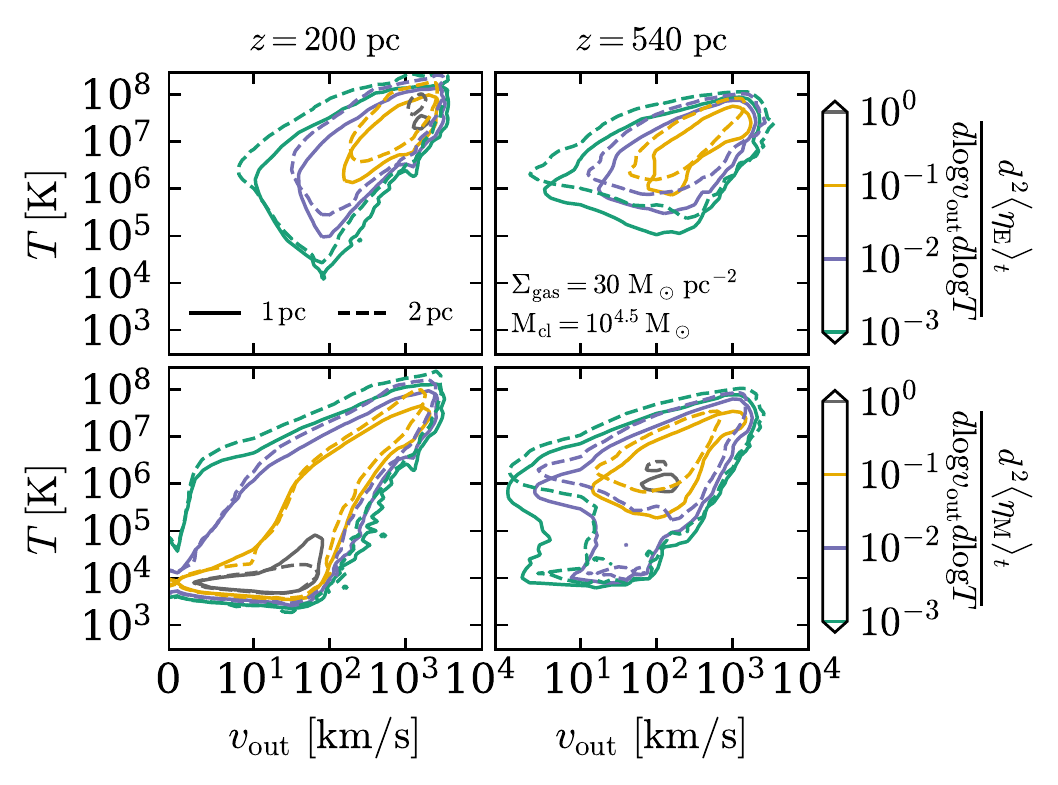}
\caption{The amount of the time averaged $\etaE$ (top row) and $\etaM$ (bottom row) per two dimensional logarithmic bin in temperature and outward velocity just above the disc at 200 pc (left column) and at 540 pc, the edge of the computational domain (right column) in $\Sigg=30\,\mparea$ $\Mcl=10^{4.5}\,\msun$ turbulent stratified simulations with $\Delta x = 1$ pc (solid contours) and 2 pc (dashed contours). At both heights the energy and mass in the higher resolution simulation is carried by systematically cooler gas due to enhanced mixing.\label{fig:detaEM_dlog_v_out_dlog_T_res}}
\end{figure}

Fig. \ref{fig:detaEM_dlog_v_out_dlog_T_res} shows the distributions of $\etaE$ and $\etaM$ in $T-\vout$ space for the two resolutions at two heights above the disc. For both quantities and at both heights the distributions are at systematically higher temperatures in the lower resolution simulation. This is due to enhanced mixing of cold and hot phases with better resolution that decreases the temperature. This enhanced mixing, however, has a minimal effect on the energy loading because the post-mixing temperature of the majority of the wind is still $T\gtrsim$ few $\times 10^6$ K where radiative cooling is inefficient. Notably, the $\etaM$ distribution in the 1 pc simulation has a tail that extends down to between $T=10^4$ and $10^5$ K and $\vout = 10$ and $100\,\kms$ that carries roughly $\etaM \sim 0.01$. This is absent in the 2 pc resolution simulation. Although this mass flux is only a small fraction of the total outflow it is indicative that with higher resolution, and/or additional physical processes there may be a larger cold component of the wind.

\section{Turbulence Realizations}\label{app:turb}

\begin{figure}
\includegraphics[width=\columnwidth]{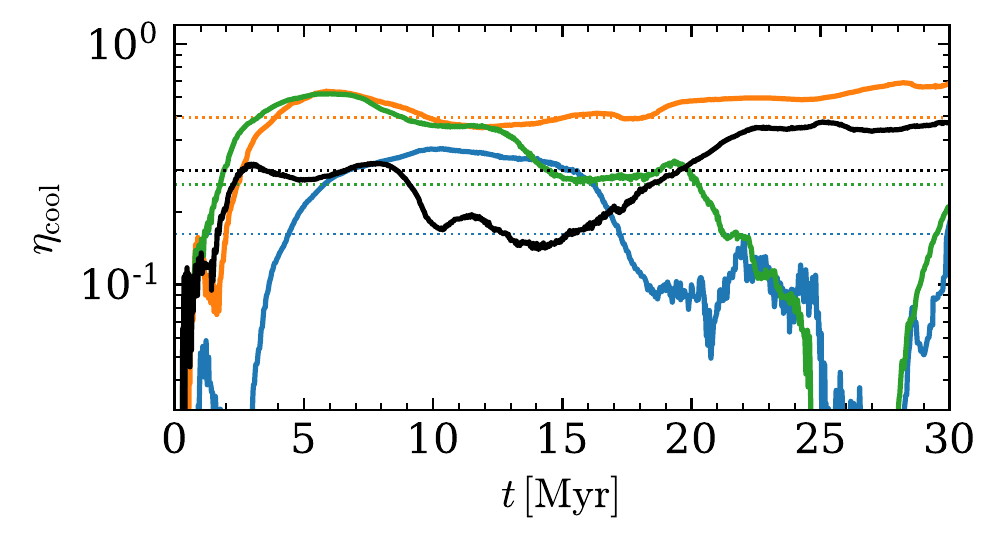}
\caption{The time evolution of $\etacool$ for four different $\Sigg=30\,\mparea$ $\Mcl=10^{4.5}\,\msun$ turbulent stratified simulations with $\Delta x = 2$ pc that differ only in turbulent driving realizations. The thin dotted lines show the time average value. The black line is the fiducial simulation discussed throughout the paper. Once the bubble breaks out in the simulation traced by the orange line the cluster efficiently vents without having to carve out a new vent. Alternatively, the clusters in the simulations traced by the blue and green simulations are impinged on by cold dense clouds that force the bubble to re-breakout. The differences in turbulent ISM properties leads to a factor of $\sim 3$ range in the average amount of cooling over the whole simulation.\label{fig:etacool_turb_dep}}
\end{figure}

The proximity of the cluster to massive cold clouds in the ISM has a large impact on the subsequent wind dynamics. These cold clouds carry a large amount of mass and momentum and when one drifts near the cluster an appreciable fraction of the cluster's energy is spent pushing and ablating the cloud. When this happens the energy remaining to power a wind is diminished. Because of this our simulations are sensitive to the properties of the turbulent ISM. Moreover, since the turbulence in our simulation is driven by hand and the location of the cluster is not tied to the local ISM properties it is likely that a given simulation may experience more or less favorable wind launching conditions (lower or higher frequency of interacting with cold clouds). To test the sensitivity of our results to the properties of the ISM we re-simulated the $\Sigg=30\,\mparea$ $\Mcl=10^{4.5}\,\msun$ turbulent stratified simulation three additional times with different random number seeds and therefore different realizations of the turbulent ISM. Fig. \ref{fig:etacool_turb_dep} shows the evolution of $\etacool$ for these four simulations. Not surprisingly there is a large degree of variability. The simulation shown with the orange line breaks out once and then is able to efficiently vent its energy virtually unimpeded for the duration of $\tsn$. On the other hand, the clusters in the simulations shown with the blue and green lines have far more difficulty keeping a channel clear for efficient venting and are nearly completely bottled up for several Myr around 20 Myr. The black line shows the fiducial simulation which lies in the middle of the range. The time average $\etacool$ ranges from 0.15 to 0.5. This factor of $\sim3$ spread in the cooling loses and accordingly the wind energy loading points to the sensitivity of wind launching to ISM properties. Simulations with self-consistently driven ISM turbulence and star formation tied to the ISM density/velocity field (e.g., gravitational collapse) will be able to assess if this degree of wind strength variation is intrinsic to the problem or a result of our idealized setup. Nevertheless, our primary finding that post-breakout cooling saps much less of the energy injected by SNe is valid regardless of the sensitivity to turbulent properties, although a given cluster may have to breakout more than once over its lifetime if ISM flows conspire to refill the cavity it excavated.

\label{lastpage}

\end{document}